\documentclass[sigconf,9pt]{acmart}

\usepackage{amsmath} 
\usepackage{textcomp} 
\usepackage{url}

\usepackage{subcaption} 
\usepackage{multirow} 
\usepackage{makecell} 
\usepackage{enumitem} 
\usepackage{soul} 

\usepackage{xcolor,listings} 
\usepackage{graphicx}     
\usepackage{booktabs}     
\usepackage{textcomp} 
\usepackage[ruled,vlined]{algorithm2e} 

\newcommand{\set}[1]{\left\{#1\right\}}
\newcommand{\parentheses}[1]{\left(#1\right)}

\newcommand\boldskipper[0]{\smallskip} 

\newcommand\jl[1]{\textcolor{red}{}}
\newcommand\minlan[1]{\textcolor{blue}{}}
\newcommand\MM[1]{\textcolor{green}{}}
\newcommand\ga[1]{\textcolor{brown}{}}
\newcommand\ran[1]{\textcolor{orange}{}}
\newcommand\sivaram[1]{\textcolor{violet}{}}
\newcommand\go[1]{\textcolor{yellow}{}}

\newcommand{\plusplus}{\hspace{-.05in}\mathrel{++}}

\setcopyright{none}
\renewcommand\footnotetextcopyrightpermission[1]{} 
\acmDOI{}
\acmISBN{}
\acmPrice{}
\acmConference{Preprint}{}{}

\begin{document}
    \title{Direct Telemetry Access}

    \author{Jonatan Langlet}
    \orcid{0000-0003-0644-6612}
    \affiliation{
       \institution{Queen Mary University of London}
    }
    \email{j.langlet@qmul.ac.uk}

    \author{Ran Ben Basat}
    \orcid{0000-0003-0196-9190}
    \affiliation{
       \institution{University College London}
    }
    \email{r.benbasat@ucl.ac.uk}

    \author{Gabriele Oliaro}
    \orcid{0000-0001-5406-0736}
    \affiliation{
       \institution{Carnegie Mellon University}
    }
    \email{goliaro@cs.cmu.edu}

    \author{Michael Mitzenmacher}
    \orcid{0000-0001-5430-5457}
    \affiliation{
       \institution{Harvard University}
    }
    \email{michaelm@eecs.harvard.edu}

    \author{Minlan Yu}
    \orcid{0000-0002-2381-0212}
    \affiliation{
       \institution{Harvard University}
    }
    \email{minlanyu@g.harvard.edu}

    \author{Gianni Antichi}
    \orcid{0000-0002-6063-4975}
    \affiliation{
       \institution{Politecnico di Milano}
       \institution{Queen Mary University of London}
    }
    \email{gianni.antichi@polimi.it}

    \renewcommand{\shortauthors}{Langlet et al.}  
    
    \begin{abstract}
        Fine-grained network telemetry is becoming a modern datacenter standard and is the basis of essential applications such as congestion control, load balancing, and advanced troubleshooting.
As network size increases and telemetry gets more fine-grained, there is a tremendous growth in the amount of data needed to be reported from switches to collectors to enable network-wide view. As a consequence, it is \mbox{progressively hard to scale data collection systems.}

We introduce Direct Telemetry Access (DTA), a solution optimized for aggregating and moving hundreds of millions of reports per second from switches into queryable data structures in collectors’ memory. DTA is lightweight and it is able to greatly reduce overheads at collectors.
DTA is built on top of RDMA, and we propose novel and expressive reporting primitives to allow easy integration with existing state-of-the-art telemetry mechanisms such as INT or Marple.

We show that DTA significantly improves telemetry collection rates. For example, when used with INT, it can collect and aggregate over $400$M reports per second with a single server, improving over the Atomic MultiLog by up to $16$x.

    \end{abstract}
    
    \maketitle

\vspace{-1.62mm}
    \section{Introduction}
\vspace{-1mm}    
    In modern data centers, telemetry is the foundation for many network management tasks such as traffic engineering, performance diagnosis, and attack detection~\cite{ben2020pint,huang2020omnimon,intelDeepInsight,kim2015band,van2017towards,yu2019network,zhou2020hypersight,zhou2020flow}. With the rise of programmable switches~\cite{trident,intelTofino1,mellanox_spectrum}, telemetry systems can now monitor network traffic in real time and at a fine granularity~\cite{INTSpec,narayana2017language,zhu2015packet, li2016flowradar,yang2018elastic,ben2020pint}. They are effectively the key enabler to support automated network control~\cite{li2019hpcc,alizadeh2014conga,heller2010elastictree} and detailed troubleshooting~\cite{guo2015pingmesh,intelDeepInsight,tammana2018distributed}. To provide network-wide views, telemetry systems also aggregate per-switch data into a centralized collector~\cite{khandelwal2019confluo, gupta2018sonata,aristaTelemetry,ciscoMDT,huaweiTelemetry,intelDeepInsight,juniperTelemetry}, commonly located in a ordinary rack within the datacenter fabric~\cite{mizrahi2018network,HuaweiTelemetry2}.

    Unfortunately, as telemetry gets more fine-grained, the amount of data to send to a collector increases and it is progressively harder to scale data collection systems~\cite{khandelwal2019confluo,van2018intcollector,zhou2020flow}. Indeed, a switch can generate up to millions of telemetry reports per second~\cite{zhou2020flow,narayana2017language} and a data center network can comprise thousands of them~\cite{guo2015pingmesh}. Also, the amount of data keeps growing with larger networks \mbox{and higher line rates~\cite{jupiter-rising-google}.} 

    Existing research boosts scalability in data collection by improving the collector's network stacks~\cite{khandelwal2019confluo,van2018intcollector}, by aggregating and filtering data at switches~\cite{intelINT,elastictrie,vestin2019programmable,zhou2020flow,narayana2017language}, or by reducing the exported information through switch cooperation~\cite{li2020concerto}.
    However, as we show, a collector can easily become either CPU- or memory bounded (\S\ref{sec:motivation}). This is due to the amount of data processing (i.e., I/O, parsing, and data insertion) it is \mbox{required to perform for every incoming report. }

    \begin{figure}[t]
        \centering
        \includegraphics[width=\columnwidth]{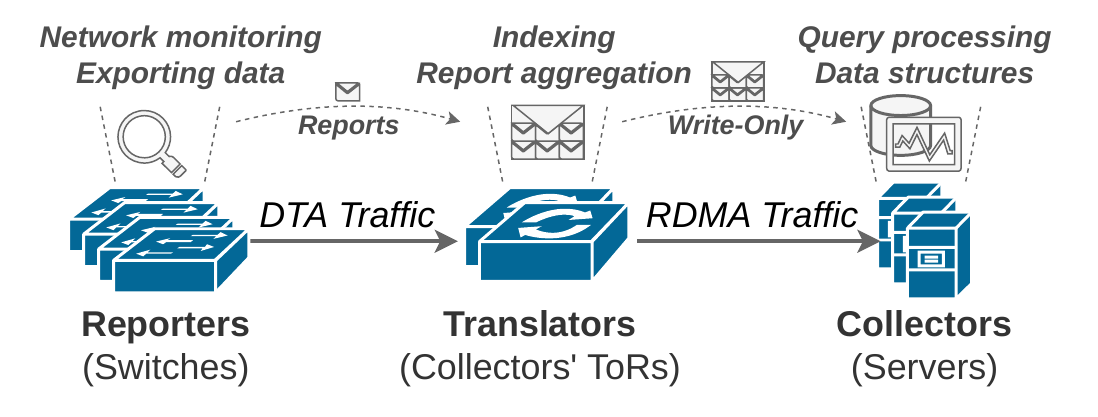}
        \caption{An overview of the telemetry data flow in DTA.}
        \vspace{-0.11009876510985123in}
        \label{fig:dta_architecture_overview}
    \end{figure}
    
    We propose \emph{Direct Telemetry Access} (DTA) --- a telemetry collection system (Figure~\ref{fig:dta_architecture_overview}) optimized for aggregating and moving hundreds of millions of reports per second from switches into queryable data structures in collectors’ memory.
    In designing DTA, we considered four key goals: (1) relieving a collector's CPU from processing incoming reports while also (2) greatly lowering the number of memory access into it. Those aspects dramatically reduce overheads at the collectors. Furthermore, we wanted (3) to be compatible with state-of-the-art telemetry reporting solutions (e.g., INT~\cite{intelINT}, Marple~\cite{narayana2017language}) while (4) imposing minimal hardware resource overheads at switches.
    
    To meet the first goal, we could simply have switches generate RDMA (Remote Direct Memory Access)~\cite{rdma} calls to a collector's memory. RDMA is available on many commodity network cards~\cite{intelE810,mellanoxConnectx6,xilinxERNIC} and can perform hundreds of millions of memory writes per second~\cite{mellanoxConnectx6}, significantly faster than the most performant CPU-based telemetry collector~\cite{khandelwal2019confluo}. Previous work~\cite{kim2020tea} has shown that one can generate RDMA instructions between a switch and a server for network functions. However, it is challenging to adopt RDMA between multiple switches and a collector for telemetry systems as RDMA performance degrades substantially when multiple clients write to the same server~\cite{kalia2016design}. Furthermore, managing RDMA connections at switches is costly in terms of hardware resources and this would conflict with our fourth goal.
    
    We instead developed a solution where the telemetry data exported by switches is encapsulated into our custom and lightweight protocol. This encapsulated data is then intercepted by the last hop switch in front of the collector (generally the Top-of-Rack switch), which we call a DTA \emph{translator}, and converted into standard RDMA calls for the corresponding memory (\S\ref{sec:overview}). 
    For the first goal, the CPU avoids processing reports by design as data is inserted into a collector's memory via RDMA.  
    For the second, the translator aggregates and batches reports before invoking RDMA calls and inserts the data in a collector's memory using RDMA-compatible write-only data structures that enable indexing of aggregates without reading from memory, thus reducing memory pressure on a collector's memory. 
    For the third, we designed several switch-level RDMA-extension primitives (\emph{Key-Write}, \emph{Postcarding}, \emph{Append}, and \emph{Key-Increment}), available to reporting switches. These are converted by the translator into standard RDMA calls and allow compatibility with many telemetry systems (\S\ref{sec:design}). 
    Finally, for the fourth, telemetry-reporting switches use our UDP-based protocol to send reports, thus freeing them from the burden of managing RDMA, which is the duty of only the translator.
    
    We implemented DTA using commodity RDMA NICs and programmable switches (\S\ref{sec:impl}) and our evaluation~(\S\ref{sec:eval}) shows that we process and aggregate over $400$M INT reports per second, without any CPU involvement, which is $16$x faster than the state-of-the-art CPU-based collector for high-speed networks~\cite{khandelwal2019confluo}. Further, when the received data can be recorded sequentially, as in the case of temporally ordered event reports, we can ingest up to a billion reports per second, $41$x more than state-of-the-art.

    \boldskipper
    \textbf{Our main contributions are:}
    \begin{itemize}[itemsep=5pt]
    \item We show that collectors can easily become either CPU- or memory bounded and this greatly limits their ability to process reports and store them in queryable data structures.
    \item We propose \emph{Direct Telemetry Access}, a novel telemetry collection system generic enough to support major telemetry reporting solutions proposed by the research community (e.g., Marple) or industry (e.g., INT).
    \item We implement DTA using commodity RDMA NICs and programmable switches.
    \item We release DTA as open source to foster reproducibility~\cite{DTARepository}.
    \end{itemize}

    \section{Motivation}
\label{sec:motivation}

\begin{table}[]
    \centering
    \resizebox{\columnwidth}{!}{
    \begin{tabular}{@{}lc@{}}
        \toprule
        \textbf{System} & \textbf{Per-switch Report Rate} \\ \midrule
        \textbf{INT Postcards} (Per-hop latency, 0.5\% sampling) & 19 Mpps\\
        \textbf{Marple~\cite{narayana2017language}} (Flowlet sizes) & 7.2 Mpps\\
        \textbf{Marple~\cite{narayana2017language}} (TCP out-of-sequence) & 6.7 Mpps\\
        \textbf{NetSeer~\cite{zhou2020flow}} (Loss events) & 950 Kpps\\\bottomrule
    \end{tabular}
    }
    \caption{Per-reporter data generation rates by various monitoring systems, as presented in their individual papers and verified through our experiments. Numbers are based on 6.4Tbps switches.}
    \vspace{-0.3in}
    \label{tab:motivation_generation_speeds}
\end{table}

Telemetry systems are commonly composed of two main components: (1) switches reporting data and (2) collectors, specialized software installed in dedicated servers located in ordinary racks within the datacenter fabric, that store the reported data~\cite{mizrahi2018network,HuaweiTelemetry2}.
As telemetry systems move to fine-grained real-time analysis with support for network-wide queries, report collection becomes \mbox{the new key bottleneck~\cite{khandelwal2019confluo}.}

We investigated several state-of-the-art telemetry systems and summarize the reporting rate generated by a single switch in Table~\ref{tab:motivation_generation_speeds}, based on the numbers available in the corresponding papers.\footnote{INT does not advertise a telemetry reporting rate. Thus, we chose an arbitrary sampling rate of 0.5\% to keep overheads reasonably low, as an example.}
For example, INT Postcard~\cite{intelINT} could generate up to $19M$ reports per second when enabled on a commodity 6.4Tbps switch and in the presence of a standard load of ${\approx}$40\%~\cite{microburstfb}. Other solutions export less data, either because they pre-process and filter data at switches~\cite{zhou2020flow,gupta2018sonata}, or because they focus on more specific tasks, \mbox{thus limiting the data to report~\cite{narayana2017language}.}

The main takeaway is that state-of-the-art solutions can easily generate \emph{millions of reports per second per switch}. 
However, to be able to gather a network-wide view at datacenter scale, we may need to collect data from thousands of switches~\cite{guo2015pingmesh} and this requires high-performance collection stacks~\cite{khandelwal2019confluo,van2018intcollector}. 
For each report from a switch, collectors spend CPU cycles to receive the data (i.e., I/O), parse it (extract content from the report), and to insert it in a queryable data structure for later use (i.e., indexing)~\cite{khandelwal2019confluo,MicrosoftTelemetryReporting,CiscoInfluxDBCollection,van2018intcollector}. 

Here, an important trade-off must be taken into account: the more complex the indexing mechanism used for final storage, the more they are suited to efficiently answer different types of queries, but in turn this generally means more CPU cycles spent in inserting data. As an example, consider a simple collector that uses only a hash table to record incoming reports. This solution is good for storing and retrieving counters (e.g., Netflow flow records~\cite{ciscoNetflow}). 
However, such a solution might be impractical for essential queries that look, e.g., at a time interval (such as analyzing losses~\cite{zhou2020flow}, congestion~\cite{INTSpec}, suspicious flows~\cite{elastictrie} or latency spikes~\cite{microburstfb} that happens at a certain point in time).
        \begin{figure}[t]
            \centering
            \begin{minipage}{0.64\columnwidth}
                \begin{subfigure}[t]{1\linewidth}
                    \centering \includegraphics[width=\linewidth]{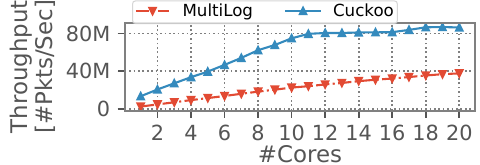}
                    \vspace{-0.1725in}
                    \caption{Collection speed.}
                    \label{fig:motiv_collector_speeds}
                \end{subfigure}
                \begin{subfigure}[t]{1\linewidth}
                    \centering \includegraphics[width=\linewidth]{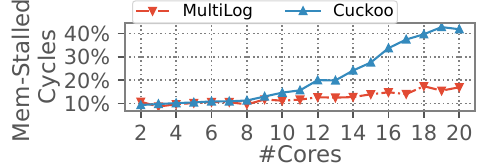}
                    \caption{Memory stalling.}
                    \label{fig:motiv_collector_stalls}
                \end{subfigure}
            \end{minipage}
            \begin{minipage}{0.35\linewidth}
                \begin{subfigure}[t]{1\linewidth}
                    \centering
                    \includegraphics[width=\linewidth]{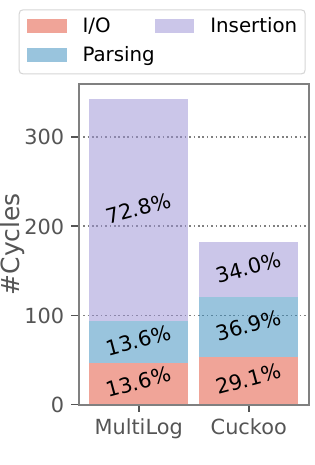}
                    \caption{Breakdown.}
                    \label{fig:motiv_collector_breakdown}
                \end{subfigure}
            \end{minipage}
            \vspace{-0.175in}
            \caption{The performance of CPU-based collectors. MultiLog is CPU bounded, while Cuckoo is memory bounded as with 20 cores, 42\% of the cycles are spent in waiting for a memory operation to finish.}
            \label{fig:motiv_cpu_collector_performance}
            \vspace{-0.1820125in}
        \end{figure}

\begin{figure}{}
            \centering
            \includegraphics[width=1.00\columnwidth]{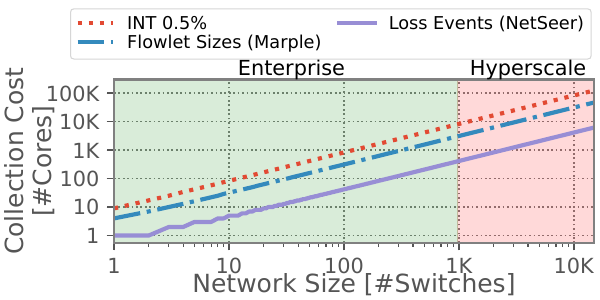}
            \caption{Number of cores needed for single-metric collection with MultiLog \mbox{at various network sizes.}}
            \label{fig:motiv_network_collection_cores}
            \vspace{-0.20in}
        \end{figure}

To better understand this trade-off, we have deployed the state-of-the-art DPDK-based telemetry collector allowing storage and diverse queries through an Atomic MultiLog~\cite{khandelwal2019confluo} (from now on we refer to it as MultiLog).\footnote{Atomic MultiLog is the basic storage abstraction in Confluo~\cite{khandelwal2019confluo}, and it is similar in interface to database tables.} 
We used a high-speed server equipped with 2x Intel Xeon Silver 4114 CPUs with 10 cores each clocked to 2.20GHz, and 2x32GB DRAM clocked to 2.67GHz. 
We compared the performance of this system to a DPDK-based lightweight solution which employs only a simple cuckoo hash table to store the received information (we refer to it as Cuckoo). 
We analyzed their behavior when receiving and storing INT reports and found that the MultiLog collector is \emph{CPU bounded}: indeed, its ability to ingest reports grows linearly with its number of cores (Figure~\ref{fig:motiv_collector_speeds}). 
Moreover, the majority of its CPU cycles, around 72.8\%, are spent in inserting the data into its internal database (Figure~\ref{fig:motiv_collector_breakdown}). 
The main takeaway is that a complex indexing scheme can have a huge toll on the performance of the collector. 
To put this in perspective, in Figure~\ref{fig:motiv_network_collection_cores}, we show the number of cores that would be needed for a growing size of a datacenter network when employing the MultiLog collector in the presence of switches reporting different information. 
Here, we can see that for networks comprising around a thousand switches~\cite{guo2015pingmesh}, we would need to dedicate nearly $10K$ cores just for collection.
For example, in a $K=28$ fat tree, this would correspond to over 11\% of the servers (assuming 16 cores each), and the problem gets worse for smaller networks.

In contrast, the lightweight Cuckoo scheme can ingest more reports per second (Figure~\ref{fig:motiv_collector_speeds}) using the same number of cores.
However, a new bottleneck arises: in our tests we see that with more than 11 cores it becomes \emph{memory bounded}.  
For example, with 20 cores, 42\% of the cycles are spent waiting for a memory operation to finish (Figure~\ref{fig:motiv_collector_stalls}). This is because the high number of reports received impose a huge stress on the memory subsystem, which needs to be read and written to parse the reports, \mbox{calculate the hashes, and resolve eventual collisions.}

Based on the observations, we enumerate our desired goals for a telemetry collection system:

\vspace{-.651mm}
\boldskipper
\noindent\textbf{Goals.} We wish to have a solution that (1) reduces as much as possible the number of cores required for data collection; (2) lowers the number of memory accesses per report; (3) is compatible with state-of-the-art telemetry reporting systems such as INT and Marple; (4) uses minimal hardware resources to get reports from switches to the collector.


    \begin{table*}[]
        \centering
        \vspace*{-2mm}
        \resizebox{\textwidth}{!}{
        \begin{tabular}{@{}lll@{}}
            \toprule
            \textbf{Primitive} & \multirow{2}{*}{\textbf{Example monitoring}} & \multirow{2}{*}{\textbf{Description}} \\
            \textit{(Interface)} & & \\ \midrule
            \multirow{5}{*}{\shortstack[l]{\textbf{Key-Write}\\\textit{(key,data)}}}
                 & INT-MD (Path Tracing) \cite{INTSpec,kim2015band} & INT sinks reporting \emph{5x4B} switch IDs using \emph{flow 5-tuple} keys \\
                 & Marple (Host counters) \cite{narayana2017language} & Reporting \emph{4B} counters using \emph{source IP} keys, through non-merging aggregation \\
                 & PacketScope (Flow troubleshooting) \cite{teixeira2020packetscope} & Report \emph{fixed-size} per-flow per-switch traversal information using \emph{<switchID,5-tuple>} as key \\
                 & PINT (Per-flow queries) \cite{ben2020pint} & \emph{1B} reports with \emph{5-tuple} keys, using redundancies for data compression through $n=f(pktID)$ \\
                 & Sonata (Per-query results) \cite{gupta2018sonata} & Reporting \emph{fixed-size} network query results using \emph{queryID} keys \\
            \hline
            \multirow{2}{*}{\shortstack[l]{\textbf{Postcarding}\\\textit{(key,hop,data)}}}
                & INT-XD/MX (Path Measurements) \cite{INTSpec,kim2015band} & Switches report \emph{4B} INT postcards using \emph{(flow 5-tuple, hop)} keys\\
                & Trajectory Sampling (Path Frequencies) \cite{duffield2001trajectory} & Collection of unique packet labels from all hops for sampled packets\\
            \hline
            \multirow{6}{*}{\shortstack[l]{\textbf{Append}\\\textit{(listID,data)}}}
                & dShark (Parser-Grouper transfer) \cite{fonseca2019dshark} & Parsers append packet summaries to lists hosted by Grouper-servers\\
                & INT (Congestion events) \cite{INTSpec,kim2015band} & INT sinks append \emph{4B} reports to a list of network congestion events \\
                & Marple (Lossy connections) \cite{narayana2017language} & Report \emph{13B} flows to a list with packet loss rate greater than threshold \\
                & NetSeer (Loss events) \cite{zhou2020flow} & Appending \emph{18B} loss event reports into network-wide list of packet losses \\
                & PacketScope (Pipeline-loss insight) \cite{teixeira2020packetscope} & On packet drop: send \emph{14B} pipeline-traversal information to central list of pipeline-loss events \\
                & Sonata (Raw data transfer) \cite{gupta2018sonata} & Appending \emph{query-specific} packet tuples from switches to lists at streaming processors\\
            \hline
            \multirow{2}{*}{\shortstack[l]{\textbf{Key-Increment}\\\textit{(key,counter)}}}
                & Marple (Host counters) \cite{narayana2017language} & Reporting \emph{4B} counters using \emph{source IP} keys, through addition-based aggregation \\
                & TurboFlow (Per-flow counters) \cite{sonchack2018turboflow} & Sending \emph{4B} counters from evicted microflow-records for aggregation using \emph{flow key} as keys \\ \bottomrule 
        \end{tabular}
        }
        \vspace{-0.0in}
        \caption{Existing telemetry monitoring systems, mapped into the primitives proposed by the current iteration of DTA.}
        \label{tab:primitive_mappings}
        \vspace{-0.3in}
    \end{table*}

    \vspace{-1mm}
    \section{Direct Telemetry Access Overview}
\label{sec:overview}
DTA leverages \emph{translators}, which are the last-hop switches adjacent to the collectors. Translators receive telemetry data from \emph{reporters} (i.e., switches exporting telemetry data), encapsulated in our lightweight custom protocol. They then aggregate and batch the reports and use standard RDMA calls to write them directly into queryable data structures in the collectors' memory (Figure~\ref{fig:dta_architecture_overview}). In the following, we discuss how, \mbox{with this architecture, we meet the goals set above.}

\boldskipper
\noindent\textbf{Meeting goal \#1.} A strawman solution to meet the first goal could have switches write their reports directly in collectors' memory with RDMA calls. This would zero any CPU requirements at collectors by design. Although this idea appears attractive, and generating RDMA instructions directly from switches is possible~\cite{kim2020tea}, it becomes problematic when applied to telemetry collection.
Namely, It is inefficient to support multiple RDMA senders writing in the same servers~\cite{kalia2016design}. This is paramount for network telemetry, where multiple switches have to report their data to a collector. Additionally, RDMA NICs can only handle a limited number of active connections (also known as \emph{queue pairs}) at high speed. Increasing the number of queue pairs degrades RDMA performance by up to 5x~\cite{dragojevic2014farm}. This limits the total number of switches that can generate telemetry RDMA packets to a collector before performance starts degrading. Alternatively, several switches can share the same queue pair, but RDMA imposes the assumption that every packet received at the collector has a strictly sequential ID, which is impractical for a distributed network of switches. 
%
DTA overcomes these challenges by having the translator, which is the last-hop switch before the collector, act as the RDMA writer.
Further, by aggregating the reports we can optimize the number of CPU cycles needed for querying as related information is stored contiguously.

\boldskipper
\noindent\textbf{Meeting goal \#2.}
We propose two techniques to lower the number of accesses into collectors' memory. 
First, we aggregate reports at the translator, thereby writing each aggregate using a single write rather than one per report. 
Second, while telemetry data has to be stored in the collectors' memory in such a way that it is easy to query~\cite{khandelwal2019confluo}, even simple data structures like hash tables often require an excessive number of memory accesses, e.g., for conflict resolution. Instead, we design RDMA-compatible write-only data structures that enable the indexing of aggregates \mbox{without reading from memory.}

\boldskipper
\noindent\textbf{Meeting goal \#3.}
We propose a number of powerful primitives available at the translator that can be used by state-of-the-art telemetry reporting systems (Table~\ref{tab:primitive_mappings}). 
The primitives abstract away many common challenges (e.g., deciding where to write data to or how to leverage the small switches' memory) and allow telemetry system designers to seamlessly benefit from our optimizations (e.g., CPU and memory accesses minimization). 

\boldskipper
\noindent\textbf{Meeting goal \#4.}
In DTA, to minimize in-network hardware resources utilization, reporting switches simply use our UDP-based lightweight protocol to send reports to the translator. 
That way, we alleviate the burden of RDMA generation and aggregation in all switches but the translators.
Indeed, the standard RDMA communication protocol (RoCEv2) requires maintaining expensive per-connection metadata and generate appropriate headers \mbox{and associated checksums.}

\section{Design}
    \label{sec:design}

    \begin{figure}
        \centering
        \includegraphics[width=1.0\columnwidth]{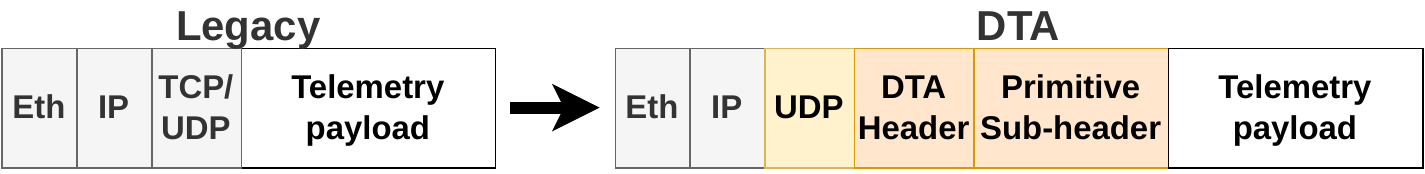}
        \vspace{-0.2in}
        \caption{DTA supports legacy telemetry systems through encapsulation with new headers.}
        \label{fig:dta_headers}
        \vspace{-0.20in}
    \end{figure}
    
    
 
 
     DTA allows easy integration with state-of-the-art telemetry monitoring systems~\cite{INTSpec,ben2020pint,narayana2017language,gupta2018sonata} through our four collection primitives that together support a wide range of telemetry solutions: \emph{Key-Write}, \emph{Postcarding}, \emph{Append}, and \emph{Key-Increment}.
     These primitives provide for placing data in the right place at the collector's memory during reporting time, so as to alleviate as much as \mbox{possible the cost of query execution.}
     
     In Table~\ref{tab:primitive_mappings}, we show that the primitives are sufficiently generic to support many state-of-the-art telemetry systems. In Figure~\ref{fig:dta_headers}, we show the structure of a DTA report. The telemetry payload exported by a switch, which depends on the specific monitoring system being used, is encapsulated into a UDP packet that carries our custom headers.
     The \emph{DTA header} (specifying the DTA primitive) and \emph{primitive sub-header} (containing the primitive parameters) are used by the translator to decide what and where to write in the collectors' data structures. This flexibility is essential as the various monitoring systems require writing telemetry in different ways for efficient analysis at the collector.
     In the following, we discuss our designed primitives. This description assumes that no DTA messages are lost, which could be either through Priority Flow Control (PFC) or a custom flow control solution as discussed later in \S\ref{sec:discussion}. The primitives themselves would still work even in case of severe in-transit loss of reports, although with degraded probabilistic guarantees which is not accounted for in the following theoretical analysis.
    
    


    \boldskipper
    \noindent\textbf{Key-Write (KW).} This primitive is designed for key-value pair collection. Storing per-flow data is one scenario where this primitive is useful (additional examples are in Table~\ref{tab:primitive_mappings}).
    
    Key-value indexing is challenging when the keys come from arbitrary domains (e.g., flow 5-tuples) and we want to map them to a small address space using just write operations.
    KW provides a probabilistic key-value storage of telemetry data and is designed for resource-efficient data plane deployments.
    We achieve this by constructing a central key-value store as a shared hash table for all telemetry-generating network switches. 
    Indexing per-key data in this hash table is performed statelessly without collaboration through global hash functions. 
    However, data written to a single memory location is highly susceptible to overwrites due to hash collisions with another key's write.
    The algorithm, therefore, inserts telemetry data as $N$ identical entries at $N$ memory locations to achieve partial collision tolerance through built-in data redundancy. In addition, a checksum of the telemetry key is stored alongside each data entry, which allows queries to be verified by validating the checksum.
    We further reduce the network and hardware resource overheads of KW by moving the indexing and redundancy generation into the DTA translator.
    This design choice effectively reduces the telemetry traffic by a factor of the level of redundancy and further reduces the telemetry report costs in the individual switches by replacing costly RDMA generation with the much more lightweight DTA protocol (\S\ref{ssec:eval_dta_vs_rdma_gencost}).
    Isolating KW logic inside collector-managing translators allows us to entirely remove this \mbox{resource cost from all other switches.}

       {
    As analyzed in Appendix~\ref{sec:kw-theory}, we can derive rigorous bounds on the probability that KW succeeds. There are two possible errors: (i) we fail to output the value for a given key; (ii) we output the wrong value for a given key. Denoting the number of slots by $M$, the number of pairs written after the queried key by $\alpha M$, and the checksum length by $b$ bits, we show that the probability of (i) is bounded by:
    {
    \addtolength{\jot}{-.212em}
    \begin{align}
        &\hspace*{-1mm}(1-e^{-\alpha\cdot N})^N \cdot (1-2^{-b})^N\label{eq1}\\
        &\hspace*{-1mm}+ (1-e^{- \alpha\cdot N})^N\cdot (1 - (1-2^{-b})^N - N\cdot 2^{-b}\cdot(1-2^{-b})^{N-1})\label{eq2}\\ 
        &\hspace*{-1mm}+ \Big( \sum_{j=1}^{N-1} {N \choose j}\cdot (1-e^{-\alpha\cdot N})^j\cdot e^{-\alpha\cdot N(N-j)}\cdot (1 - (1-2^{-b})^j) \Big).\label{eq3}
        \\[-2.5ex]\notag
    \end{align}
    }
    Here, \eqref{eq1} bounds the probability that all $N$ locations are overwritten with other checksums; \eqref{eq2} bounds the probability that all locations are overwritten and at least two items with our key's checksum write different values;
    and \eqref{eq3} bounds the probability that not all slots are overwritten, but at least one is overwritten with the query key's checksum.  
%
%
    We also bound the probability \mbox{of giving the wrong output (ii) by}
    \begin{equation}
        (1-e^{-\alpha\cdot N})^N\cdot  N\cdot 2^{-b}.\label{eq4_0}
    \end{equation}
    \addtolength{\jot}{.32em}
    For example, if $N=2, b=32, \alpha=0.1$, the chance of not providing the output is less than 3.3\%, while the probability of wrong output is bounded by $1.6\cdot 10^{-11}$. This aligns with the best effort standard of network telemetry (e.g., INT is often collected using UDP, and packet loss results in missing reports) while having a negligible chance of wrong output. Note that this error is significantly lower than with $N=1$ (which results in not providing output with probability 9.5\%) and higher than for $N=4$ (probability 1.2\%). However, increasing $N$ also has implications to throughput (more RDMA writes) and is not always justified; we elaborate on this tradeoff \mbox{in \S\ref{sub:KW-perf} and show that $N=2$ is often a good compromise.}
    }

    DTA also lets switches specify the importance of per-key telemetry data by including the level of redundancy, or the number of copies to store, as a field in the KW header.
    Higher redundancy means a longer lifetime before being overwritten, as we discuss in \S\ref{sec:eval_keywrite_redundancy_effectiveness}. As the level of redundancy used at report-time may not be known while querying, the collector can assume by default a maximum (e.g., $4$) redundancy level. If the data was reported using fewer slots, unused \mbox{slots would appear as overwritten entries (collision).} 
    
    \boldskipper
    \noindent\textbf{Postcarding.} 
    One of the most popular INT working modes is postcarding (INT-XD/MX~\cite{INTSpec}), where switches generate \emph{postcards} when processing selected packets and send them to the collector (e.g., for tracing a flow's path.) A report is a collection of one postcard from each hop.
    Intuitively, while we could use the KW primitive to write all postcards for a given packet, this is potentially inefficient for several reasons. First, each packet can trigger multiple reports that will use multiple RDMA writes even if $N=1$ (e.g., one per switch ID for path tracing.) In turn, for answering queries with KW (e.g., outputting the switch ID list), the collector will need to make multiple random-access reads, which is slow. 
    Further, adding the KW's checksum to each hop's information is wasteful and degrades the memory-queryability tradeoff. For ease of presentation, we first explain how to reduce the number of writes and \mbox{later elaborate on how to decrease the width of each slot.}
    
    \begin{figure}
        \centering
        \includegraphics[width=1.0\columnwidth]{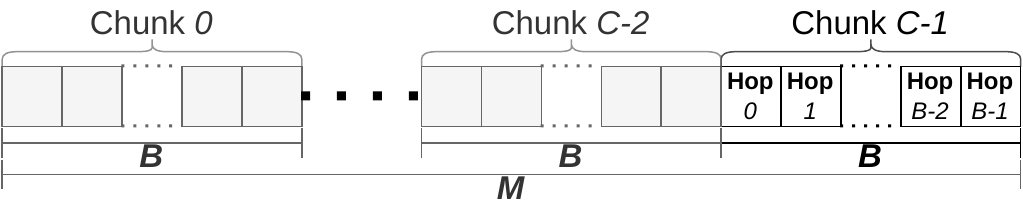}
        \vspace{-0.3in}
        \caption{{\small The Postcarding memory structure at the collector.}}
        \label{fig:postcarding_structure}
        \vspace{-0.2in}
    \end{figure}
    
    Our observation is that if we know a bound $B$ on the number of hops a packet traverses (e.g., five for fat tree topology), we can improve the above by writing all of a packet's
    postcards into a consecutive memory block.
    To that end, we break the $M$ memory locations into chunks of size $B$, yielding $C=M/B$ chunks.
    The $i$'th postcard for a packet/flow ID $x$ is written into  $B{\cdot} h(x)+i$, where $h$ maps identifiers into chunks (i.e., $h(x)\in \set{0,\ldots,C-1}$). 
    This way, the report for all up to $B$ is \mbox{consecutive in the memory, as shown in Figure~\ref{fig:postcarding_structure}.} 

    To reduce the number of RDMA writes, we use a mapping from IDs to postcards at the translator. That is, the translator shall cache postcards $0,1,\ldots,B-1$ before writing the report to the collector's memory using a single RDMA write, once $B$ flow postcards are counted in the translator. Further, answering queries will thus require a single memory random access.
    As not all packets follow a $B$ hop path, egress switches can provide a packet's path length inside postcards, and translators can use this value to trigger writes before the postcard-counter reaches $B$. Additionally, reports may be flushed early due to collisions on the switch cache. 


    Finally, we reduce the number of bits needed for each location, compared with writing the value and checksum to each slot. Intuitively, we leverage the $B$ postcards to amplify the success probability -- we output the report only if \emph{all} checksums are valid thereby minimizing the chance of wrong outputs. 
    Intuitively, we use $b > \log_2 |V|$ bits per location to get a collision chance of $\approx(|V|\cdot 2^{-b})$ for each location and $\approx(|V|\cdot 2^{-b})^B$ overall.  Here, $V$ is the set of all possible values (e.g., all switch IDs).
    As noted by PINT~\cite{ben2020pint}, $|V|$ is often smaller than $2^{32}$ (although the INT standard requires that each value is reported using exactly four bytes~\cite{INTSpec}), allowing us to use small $b$ values.
    
    Let  $g$ be a hash function that maps values $v\in V$ into $b$-bit bitstrings, where $b$ is the desired slot width. We use a ``blank'' value $\sqcup$ to denote that values for a given hop were not collected (potentially because the path length was shorter than $B$); this way, each flow always writes all $B$ hops' values, minimizing the chance of false output due to hash collisions.
    Then, when receiving a postcard value $v_{x,i}\in V$ from the $i$'th hop of flow/packet ID $x$, we write $\mbox{checksum}(x,i)\oplus g(v_{x,i})$ into location $B\cdot h(x)+i$ (here $\mbox{checksum}(x,i)$ also returns a $b$-bit result and $\oplus$ is the bitwise-xor operator). When answering queries about $x$, we check if there exists $\ell$ such that for all $i\in\set{0,\ldots,\ell-1}$ there exists a value $v_{x,i}\in V$ for which $\mbox{checksum}(x,i)\oplus g(v_{x,i})$ is stored in slot $B\cdot h(x)+i$ and for all $i\in\set{\ell,\ldots,B-1}$ $\mbox{checksum}(x,i)\oplus g(\sqcup)$ is stored. If so, we output that the postcard reports were $v_{x,0},v_{x,1},\ldots,v_{x,\ell-1}$. In this case, we say that the chunk contains valid information.
    We further note that checking the existence of such $v_{x,i}$ can be done in constant time using a pre-populated lookup table that stores all key-value pairs $\set{(g(v),v)\mid v\in V\cup\set{\sqcup}}$.
    
    
    Our approach generalizes with redundancy $N>1$: we use $N$ hash functions $h_1,\ldots,h_N$ such that $\mbox{checksum}(x,i)\oplus g(v_{x,i})$ is written into locations $\set{B\cdot h_j(x)+i\mid j\in\set{1,\ldots,N}}$. For answering queries, we output $v_{x,0},v_{x,1},\ldots,v_{x,\ell-1}$ if it appears in a valid subset of the $N$ chunks and all other chunks contain invalid information.
    \allowdisplaybreaks

\vspace{1mm}
    In Appendix~\ref{app:postcarding}, we analyze the primitive and prove that the probability of not providing an output is bounded by:
    {\small
    \begin{align}
        &\ (1-e^{-\alpha\cdot N})^N\cdot \Big({1-\parentheses{(|V|+1)\cdot2^{-b}}^B\Big)}^{N}\label{eq5a}\\
        +&\ (1-e^{-\alpha\cdot N})^N\cdot\bigg(1-\parentheses{1-\parentheses{(|V|+1)\cdot2^{-b}}^B}^{N}\notag\\&\ - N\cdot \parentheses{(|V|+1)\cdot2^{-b}}^B\cdot\parentheses{1-\parentheses{(|V|+1)\cdot2^{-b}}^B}^{N-1}\bigg)\label{eq6a}\\
        +&\ \sum_{j=1}^{N-1} {N \choose j}\cdot (1-e^{-\alpha\cdot N})^j\cdot e^{-\alpha\cdot N(N-j)}\notag\\[-2ex]&\qquad\qquad\qquad\cdot \bigg({1-\parentheses{1-\parentheses{(|V|+1)\cdot2^{-b}}^B}^{j}}\bigg).\label{eq7a}
    \end{align}
    }
    We also show that the chance of wrong output is bounded by:\vspace*{-1.04mm}
    \begin{align}
        (1-e^{-\alpha\cdot N})^N\cdot  N\cdot \parentheses{ (|V|+1)\cdot2^{-b}}^ B
        .\label{eq8a}
        \\[-4.5ex]\notag
    \end{align}\vspace*{-2mm}
    \addtolength{\jot}{.32em}
    \looseness=-1
    
    We consider a numeric example to contrast these results with using KW for each report of a given packet. Specifically, suppose that we are in a large data center ($|V|=2^{18}$ switches) and want to run path tracing by collecting all (up to $B=5$) switch IDs using $N=2$ redundancy.
    Further, let us set $b=32$-bit per report and compare it with $64$ bits ($32$ for the key's checksum and $32$ bits for the switch ID) used in KW, and that $C\cdot \alpha$ packets' reports were collected after the queried one, for $\alpha=0.1$.
    The probability of not outputting a collected report (\ref{eq5a}-\ref{eq7a}) is then at most 3.3\% and the chance of providing the wrong output \eqref{eq8a} is lower than $10^{-22}$. In contrast, using KW for postcarding gives a false output probability of $\approx 8\cdot 10^{-11}$ (in at least one hop) \mbox{using twice the bit-width per entry!}
    
    \boldskipper
    \noindent\textbf{Append.} Some telemetry scenarios are not easily managed with key-value stores.
    A classic example is when a switch exports a stream of events, where a report would include an event identifier and an associated timestamp
    (e.g., packet losses~\cite{zhou2020flow}, congestion events~\cite{INTSpec}, suspicious flows~\cite{elastictrie}, latency spikes~\cite{microburstfb}). 
    We thus provide a primitive that allows reporters to append information into global lists, with a pre-defined telemetry category in each list.
            
    Telemetry reporters simply have to craft a single DTA packet declaring what data they want to append to which list, and forward it to the appropriate collector. The translator then intercepts the packet and generates an RDMA call to insert the data in the correct slot in the pre-allocated list. The translator utilizes a pointer to keep track of the current write location for each list, allowing it to insert incoming data per-list. Append adds reports sequentially and contiguously into memory. This leads to an efficient use of memory and strong query performance. Translation also allows us to \emph{significantly} improve on the collection speeds by batching multiple reports together in a single RDMA operation.

    \boldskipper
    \noindent\textbf{Key-Increment (KI).} 
    This is similar to the KW primitive, but allows for addition-based data aggregation. That is, the KI primitive does not instruct the collector to set a key to a specific value, but it instead \emph{increments} the value of a key. For example, switches might only store a few counters in a local cache, and evict old counters from the cache periodically when new counters take their place~\cite{sonchack2018turboflow, narayana2017language}. The KI primitive can then deliver collection of these evicted counters at RDMA rates. As with KW, the translator reduces network overheads compared with a more naive design.
            
    Our KI memory acts as a Count-Min Sketch~\cite{cormode2005improved} and we increment $N$ values using the RDMA Fetch-and-Add primitive. On a query, KI returns the minimum value from these $N$ locations. Hash collisions may lead to an overestimate of the value, with error guarantees matching those of Count-Min Sketches~\cite{cormode2005improved}. 
    The counters' memory \mbox{may be reset periodically, depending on the application.}
        
    \boldskipper
    \noindent\textbf{Extensibility.} DTA is easily extensible to other primitives by the addition of new translation paths at translators, although they would remain constrained by the limitations imposed by commodity programmable switches~\cite{miao2017silkroad}.  Some of these limitations could be overcome by implementing the translator logic into FPGA-based smartNICs (see \S\ref{sec:discussion}).
    For example, one could extend DTA to support collection of sketch-based measurements. This could allow for either in-network discovery of network-wide heavy hitters, or aggregation of counters at the translator to decrease the collection load at compute servers.
    Additionally, the translator does not have to be a semi-passive data aggregator as presented here, and primitives could be designed to be more active. For example, one could use techniques similar to the ones presented by Gao et al.~\cite{gao2021stats} to derive the network state directly at the translator based on the intercepted telemetry reports, thereby offloading even parts of analysis from the telemetry collectors.

\section{Implementation}\label{sec:impl}
    \vspace*{1mm}

        Our codebase includes approximately $5K$ lines of code divided between the logic for the DTA reporter (\S\ref{ssec:impl-reporter}), the translator (\S\ref{ssec:impl-translator}), and collector RDMA service (\S\ref{ssec:impl-collector}).
        The hardware resource footprints are presented later in Sections \S\ref{ssec:eval_dta_vs_rdma_gencost} and \S\ref{sec:eval_asic_resources}.
        DTA is released in open-source~\cite{DTARepository}.

    \subsection{DTA Reporter}
    \label{ssec:impl-reporter}
    \vspace*{1mm}
    The reporter takes $\approx 700$ lines of P4\_16 for the Tofino ASIC. Controller functionality is written in $\approx 100$ Python lines, and is responsible for populating forwarding tables and inserting collector IP addresses for the DTA primitives.
            
    DTA reports are generated entirely in the data plane and the logic is in charge of encapsulating the telemetry report into a UDP packet followed by the two DTA-specific headers where the primitive and its configuration parameters are included.

    
    \subsection{DTA Translator}
    \label{ssec:impl-translator}
    \begin{figure*}[hbt!]
        \centering
        \includegraphics[width=0.9\linewidth]{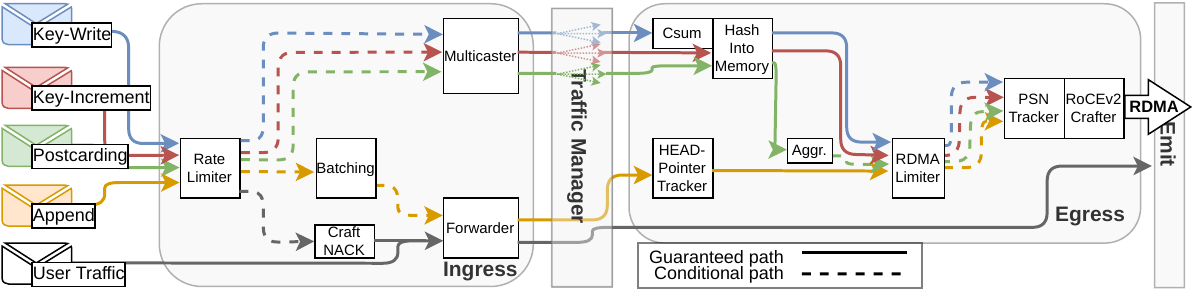}
        \vspace{-1mm}
        \caption{A translator pipeline with support for Key-Write, Key-Increment, Postcarding, and Append. Five paths exist for pipeline traversal, used to process different types of network traffic in parallel while efficiently sharing pipeline logic.}
        \label{fig:translator_pipeline}
    \end{figure*}
    The translator has a control program written in $800$ lines of Python that runs on the switch CPU.
    It is in charge of setting up the RDMA connection to the collector by crafting RDMA Communication Manager (RDMA\_CM) packets, which are then injected into the ASIC.

    The translator pipeline (Figure~\ref{fig:translator_pipeline}) is written in $2K$ lines of P4\_16 for the Tofino ASIC. 
    This pipeline includes support for internal processing of the DTA primitives, RDMA generation, basic user-traffic forwarding, as well as RDMA queue-pair resynchronization and rate limiting to ensure stable RDMA connections in case of congestion events at the collectors' NICs. Rate limiting can be configured to generate a NACK sent back to the reporter in case of a dropped report during these congestion events.

    \noindent The RDMA logic is shared by all primitives. This includes lookup tables filled with RDMA metadata, SRAM storage for the queue pair packet sequence numbers, and the task of crafting RoCEv2 headers. 
    The DTA packets themselves are used as the base for RDMA generation. This is done by completely substituting the DTA headers with the specific \mbox{RoCEv2 headers required by the DTA operation.}
    
    \noindent The redundancy in Key-Write, Key-Increment, and Postcarding is generated by the packet replication engine through multicasting (\emph{Multicaster} in Figure~\ref{fig:translator_pipeline}).
    The switch CPU crafts specific multicast rules to force the ASIC to emit several packets at the correct egress port as triggered by a single DTA ingress. 
    
    \boldskipper
    \noindent\textbf{Key-Write} and \textbf{Key-Increment} both follow the same fundamental logic, with the main difference being the RDMA operation that they trigger.
    Key-Write triggers RDMA Write operations, while a Key-Increment triggers RDMA Fetch-and-Add. Both cause $N$ packet injections into the egress pipeline, using the multicast technique. The Tofino-native CRC engine is used to calculate the $N$ memory locations, and is also used to calculate a concatenated $4B$ checksum for Key-Write.
    Carefully selected CRC polynomials are used to create several independent hash functions using \mbox{the same underlying CRC engine.}
    
    \boldskipper
    \noindent\textbf{Postcarding} uses an SRAM-based hash table with 32K slots storing fixed-size 32-bit payloads. The Tofino-native CRC engine is used for indexing and value encoding. The hop-specific checksums are implemented through custom CRC polynomials instead. Emissions are triggered either by a collision or when a row counter reaches the path length. 
    We note that an efficient implementation requires the RDMA payload sizes to be powers of 2 (due to bitshift-based multiplication during address calculation) and the chunk sizes are therefore padded from $5*4B=20B$ to $32B$, trading storage efficiency for a reduced switch footprint.
    
    \boldskipper
    \noindent\textbf{Append} has its logic split between ingress and egress, where ingress is responsible for building batches, and egress tracks per-list memory pointers. Batching of size $B$ is achieved by storing $B-1$ incoming list entries into SRAM using per-list registers. Every $Bth$ packet in a list will read all stored items, and bring these to the egress pipeline where they are sent as a single RDMA Write packet. Lists are implemented as ring-buffers, and the translator keeps a per-list head pointer to track where in server memory the next batch should be written. Our prototype supports tracking up to $131K$ simultaneous lists.

    \subsection{Collector RDMA Service}\label{ssec:impl-collector}
        The collector is written in $1.3K$ lines of C++ using standard Infiniband RDMA libraries, and has support for per-primitive memory structures and querying the reported telemetry data. The collector can host several primitives in parallel using unique RDMA\_CM ports, and advertise primitive-specific {metadata to the translator using RDMA-Send packets.}

    \begin{figure}
        \centering
        \begin{subfigure}[t]{\columnwidth}
            \centering \includegraphics[width=\linewidth]{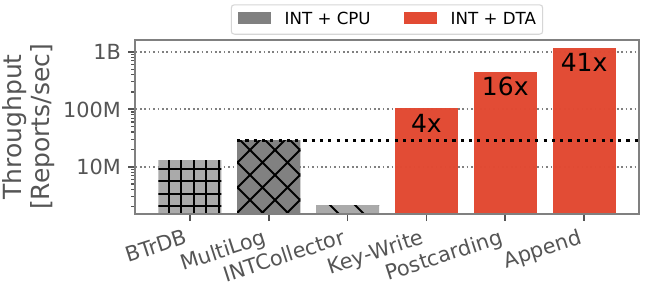}
            \vspace{-0.2in}
            \caption{Generic $4B$ INT collection.}
            \label{fig:eval_e2e_int}
            \vspace{0.2in}
        \end{subfigure}
        \\\vspace{-4mm}
        \begin{subfigure}[t]{\columnwidth}
            \centering
            \includegraphics[width=\linewidth]{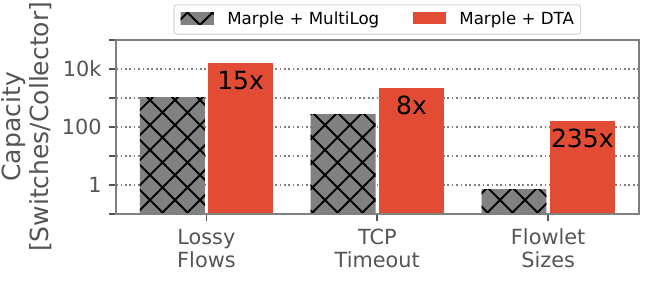}
            \vspace{-0.2in}
            \caption{Diverse Marple queries.}
            \label{fig:eval_e2e_marple}
        \end{subfigure}
        \vspace{-0.164208in}
        \caption{A performance comparison of DTA against state-of-the-art CPU-collectors. These use 16 cores for data ingestion, while DTA essentially bypasses the CPU entirely for data ingestion by using RDMA. (b) MultiLog vs DTA when using Marple as a monitoring system running on switches.}
        \label{fig:eval_e2e}
         \vspace{-4mm}
    \end{figure}

    \section{Evaluation}\label{sec:eval}
    \vspace{2mm}

    In this section, we show that:
    \begin{itemize}[align=left, leftmargin=\parindent, labelindent=.3210\parindent, listparindent=.210\parindent, labelwidth=0pt]
        \item DTA supports very high collection rates (\S\ref{sub:throughput}).
        \item DTA imposes a negligible memory pressure at collectors (\S\ref{sub:memory}).
        \item DTA is lightweight (\S\ref{ssec:eval_dta_vs_rdma_gencost}, \S\ref{sec:eval_asic_resources}).
        \item DTA's primitives are fast (\S\ref{sub:KW-perf}, \S\ref{sub:postcard-perf}, \S\ref{sec:eval_append_performance}).
    \end{itemize}
    
    We use two x86 servers connected through a BF2556X-1T~\cite{bf2556XTofino} Tofino~1~\cite{intelTofino1} switch with 100G links. Both servers mount 2x Intel Xeon Silver 4114 CPUs, 2x32GB DDR4 RAM @ 2.6GHz, and run Ubuntu 20.04 (kernel 5.4). One of them serves as a DTA report generator using TRex~\cite{ciscoTRex}. The other, equipped with an RDMA-enabled Mellanox Bluefield-2 DPU~\cite{mellanoxBluefield2}, serves as the collector. Here, server BIOS has been optimized for high-throughput RDMA~\cite{perfTuningMellanox}, and all RDMA-registered memory is allocated on 1GB huge pages.
    
    
    \subsection{DTA in Action}
    \vspace{2mm}
    \label{sub:throughput}
        We first investigate if DTA scales better than CPU-based collectors in the presence of telemetry volumes generated by large-scale networks.
        To do so, we compare the performance of DTA and state-of-the-art CPU-collectors when coupled with two different monitoring systems: INT~\cite{INTSpec,kim2015band} and Marple~\cite{narayana2017language}.
        Here, we use a DTA configuration with $N=1$ and batching of size $16$, while CPU-collectors {use $16$ dedicated CPU cores in the same NUMA-node.}
        
        The collectors in Figure~\ref{fig:eval_e2e_int} collect generic $4B$ INT reports that are available for offline queries using the flow 5-tuple as the key.
        We test INTCollector~\cite{van2018intcollector}, to the best of our knowledge the only open source INT collector that uses InfluxDB for storage.
        We also study BTrDB~\cite{andersen2016btrdb}, and the state-of-the-art solution for high-speed networks, Confluo, which is based on MultiLog technology.
        Key-Write inserts each report into its key-value store, and Postcarding assumes 5-hop aggregation with no intermediate reports. Append instead inserts the reports into one of the available data lists. 
        As Figure~\ref{fig:eval_e2e_int} shows, DTA improves on key-based INT collection by at least $4$x, or up to $16$x when aggregating the postcards into 5-hop tuples, with even higher performance gains if pre-categorized \mbox{and chronological storage through Append suffices.}
        
        \looseness=-1
        We also integrated Marple with DTA and MultiLog and configured them to support the same queries against the collected data (i.e., Lossy Flows, TCP Timeout, and Flowlet Sizes). Here, \emph{Lossy Flows} reports high loss rates together with their corresponding flow 5-tuples, and DTA uses the Append primitive to store the data chronologically in several lists, allowing operators to retrieve the most recently reported network flows with packet loss rates in one of several ranges.
        \emph{TCP Timeouts} reports the number of TCP timeouts per-flow in recent time, and DTA uses the Key-Write primitive to allow operators to query the number of timeouts experienced by any arbitrary flow.
        \emph{Flowlet Sizes} reports flow 5-tuples together with the number of packets in their most recent flowlets, and DTA appends the flow identifiers to one of the available lists to allow the construction of per-flow histograms of flowlet sizes.
        
        We experimented using real data center traffic~\cite{benson10imc} and found that DTA increases the number of Marple reporters (i.e., network switches) that a collector can support before the rate of data generation overwhelms the collector (Figure~\ref{fig:eval_e2e_marple}). 
        Their queries cost as well as their performances \mbox{are analyzed in later sections (\S\ref{sub:KW-perf}, \S\ref{sec:eval_append_performance}).}

        \nopagebreak
        \boldskipper
        \noindent\textbf{Takeaway:} DTA improves on data collection speeds compared with CPU-based collectors by \emph{one to two orders of magnitude} when integrated with state-of-the-art telemetry systems, while supporting the same types of queries.

\subsection{Reduced Memory Pressure}
\vspace{2mm}
\label{sub:memory}
    \begin{figure}
        \centering
        \includegraphics[width=1.0\columnwidth]{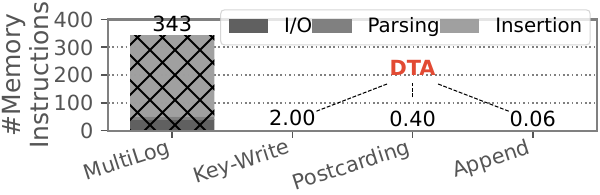}
        \caption{Average number of memory instructions per-report for ingestion of INT postcards.}
        \vspace{-0.035in}
        \label{fig:eval_mem_operations}
    \end{figure}
    In Figure~\ref{fig:eval_mem_operations}, we present the average number of memory instructions required per-report for the DTA primitives, when configured with a redundancy level of 2, path length of 5 hops, and batch size of 16 elements.
    DTA imposes a low pressure on memory. This is achieved mostly because no accesses are needed for I/O and report parsing, regardless of the indexing scheme used.
    Some DTA primitives use less than a single memory instruction per report on average, owing to its aggregation and batching techniques, which can intelligently insert several reports simultaneously with a single RDMA operation.
    For example, Key-Write, the primitive that imposes the heaviest load on memory, needs just \mbox{$0.58\%$ as many accesses as MultiLog.}

    \nopagebreak
    \boldskipper
    \noindent\textbf{Takeaway:} DTA \emph{significantly} reduces the number of memory accesses required for report ingestion.

    \subsection{Reporter Resource Footprint of DTA}
    \vspace{2mm}
    \label{ssec:eval_dta_vs_rdma_gencost}

\begin{figure}[]
        \centering
        \includegraphics[width=1.0\columnwidth]{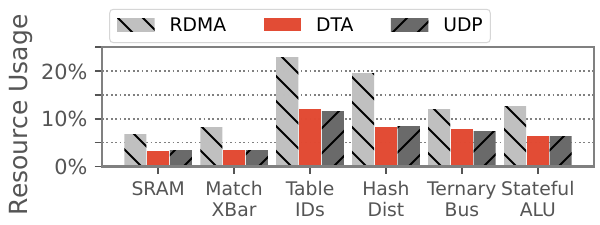}
        \vspace{-0.13in}
        \caption{Hardware resource costs of a DTA Reporter compared to an RDMA-generating reporter, and a baseline UDP-based reporter. Note how DTA {imposes an almost identical resource footprint to UDP.}}
        \label{fig:dta_vs_rdma_resource_costs}
    \end{figure}

    We compared the hardware costs associated with generating DTA reports against either directly emitting RDMA calls from switches, or creating UDP-based messages as generally done by CPU-based collectors.
    We used a switch implementing a simple INT-XD system and, in Figure~\ref{fig:dta_vs_rdma_resource_costs}, we show the cost associated with the change of its report-generation mechanism. Here, we see that DTA is as lightweight as UDP, while RDMA generation is much more expensive.

    \nopagebreak
    \boldskipper
    \noindent\textbf{Takeaway:} DTA halves the resource footprint of reporters compared with RDMA-generating alternatives, and has a \emph{similar resource footprint to simple UDP generation}. 

    \subsection{Translator Resource Footprint}\label{sec:eval_asic_resources}
    \vspace{0.105in}

    \begin{table}[]
        \centering
        \resizebox{\columnwidth}{!}{
            \begin{tabular}{@{}llllll@{}}
                \toprule
                \textbf{} & \textbf{SRAM} & \begin{tabular}[c]{@{}l@{}}\textbf{Match}\\ \textbf{Crossbar}\end{tabular} & \begin{tabular}[c]{@{}l@{}}\textbf{Table}\\ \textbf{IDs}\end{tabular} & \begin{tabular}[c]{@{}l@{}}\textbf{Ternary}\\ \textbf{Bus}\end{tabular} & \begin{tabular}[c]{@{}l@{}}\textbf{Stateful}\\ \textbf{ALU}\end{tabular} \\ \midrule
                \textbf{Base footprint} & 13.2\% & 10.6\% & 49.0\% & 30.7\% & 25.0\% \\
                \textbf{Batching} & +3.2\% & +7.2\% & +7.8\% & +7.8\% & +31.3\% \\ \bottomrule
            \end{tabular}
        }
        \vspace{0.0616in}
        \caption{Resource footprint of a translator in Tofino while supporting Key-Write, Postcarding, and Append. Append is batching 16x4B reports.}
        \vspace{-0.135in}
        \label{tab:resource_costs}
    \end{table}
    
    Table~\ref{tab:resource_costs} shows the resource usage of the translator, alongside the additional costs of including Append batching.
    The footprint of the DTA translator is mainly due to its concurrent built-in support for several different primitives. Application-dependent operators might \mbox{reduce their hardware costs by enabling fewer primitives.}
    
    Batching of Append data has a relatively high cost in terms of memory logic (Stateful ALU), due to our non-recirculating RDMA-generating pipeline requiring access to all $B-1$ entries during a single pipeline traversal. 
    It is worth noting that batching also has the potential for a tenfold increase in collection throughput, and we conclude that it is a worthwhile tradeoff. A compromise is to reduce the batch sizes, as they linearly correlate with the number of additional stateful ALU calls.
    
    Deploying multiple simultaneous Append-lists does not require additional logic in the ASIC, it just necessitates more statefulness for keeping per-list information (e.g., head-pointers and per-list batched data).
    Note that the actual SRAM footprint of the translator is small, and tests show that the translator can support hundreds of thousands of simultaneous lists for complex setups, which is much more than the $255$ lists included at the time of evaluation.
            
    

    \nopagebreak
    \boldskipper
    \noindent\textbf{Takeaway:} A translator pipeline which simultaneously supports the Key-Write, Postcarding, and Append primitives fits in first-generation programmable switches, while \emph{leaving a majority of resources freed up for other functionality}. Batching can impose a high toll on the Stateful ALUs.

    \subsection{Key-Write Primitive Performance}
    \vspace{2mm}
    \label{sub:KW-perf}
    \begin{figure}
        \centering
        \includegraphics[width=\columnwidth]{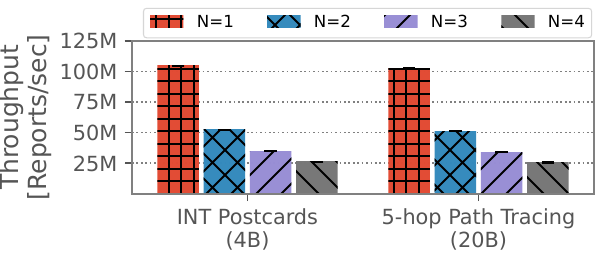}
        \vspace{-0.1523in}
        \caption{Per-flow path tracing collection rates, using the DTA Key-Write primitive, either as INT-XD/MX postcards (4B) or full 5-hops path as in INT-MD (20B).}
        \vspace{-0.06102in}
        \label{fig:eval_keywrite_rate}
    \end{figure}
                
    We have benchmarked the collection performance of the DTA Key-Write primitive using INT as a use case.
    We instantiated a $4GiB$ key-value store at the collector and had the translator receive either $4B$ or $20B$ encapsulated INT messages from the reporter (our traffic generator). 
    The former case emulates the scenario of having INT working in postcard mode with event detection (so some hops may not generate a postcard), while the latter reproduces an INT path tracing configuration on a 5-hops topology where the last hop reports data to a collector.
    We repeated the test using different levels of redundancy ($N$) and reported the results we obtained in Figure~\ref{fig:eval_keywrite_rate}. 
    Notice the expected linear relationship between the throughput and level of redundancy since each incoming report will generate $N$ RDMA packets towards the collector. However, one might still prefer the performance tradeoff against the increased data robustness in the collector storage, which allows for successful queries against much older telemetry reports.
    Furthermore, the collection rate is unaffected by the increase in the telemetry data size until the $100Gbps$ line rate is reached. In our tests, we saw that \mbox{this was the case for telemetry payloads of $16B$ or larger.} 

    \nopagebreak
    \boldskipper
    \noindent\textbf{Takeaway:} Key-Write can collect \emph{$100M$ INT reports per second} and its performance depends on the redundancy level.
    
    \subsubsection{Key-Write Query Speed} 
    
    \begin{figure}[]
        \centering
        \begin{subfigure}[]{0.577\columnwidth}
            \centering \includegraphics[width=\linewidth]{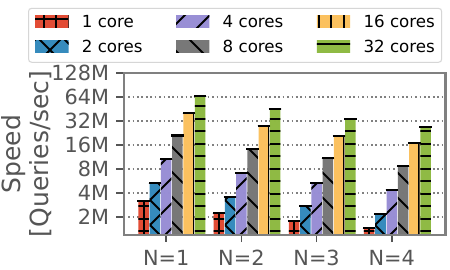}
            \caption{Query rate}
            \label{fig:eval_keywrite_querying_speed}
        \end{subfigure}
        \hfill
        \begin{subfigure}[]{0.404\columnwidth}
            \centering
            \includegraphics[width=\linewidth]{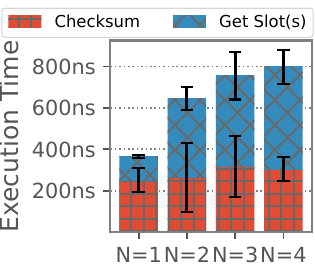}
            \caption{Per-query breakdown}
            \label{fig:eval_keywrite_querying_breakdown}
        \end{subfigure}
        \caption{Key-Write primitive querying performance.}
        \vspace{-0.02in}
        \label{fig:eval_keywrite_querying}
        \end{figure}
    
    Querying for data stored in our key-value store using the Key-Write primitive requires the calculation of several hashes.        
    Here we evaluate the \emph{worst case} performance scenario, when the collector has to retrieve every redundancy slot before being able to answer a query.
    Specifically, we queried $100M$ random telemetry keys, with a key-value data structure of size $4GiB$ containing $4B$ INT postcards data alongside $4B$ concatenated checksums for query validation. Figure~\ref{fig:eval_keywrite_querying_speed} shows the speed at which the collector can answer incoming telemetry queries using various redundancy levels ($N$). 
    
    Key-Write query processing can be easily parallelized, and we found the query performance to scale near-linearly when we allocated more cores for processing. For example, 4 cores could query $7.1$ million flow paths per second with $N=2$, while 8 cores manage $14.2$ million queries per second.
    
    Figure~\ref{fig:eval_keywrite_querying_breakdown} shows the time breakdown serving queries. Most of the execution time is spent calculating CRC hashes, for either verifying the concatenated checksum (\emph{Checksum}), or calculating {memory addresses of the $N$ redundancy entries (\emph{Get Slot}).}
    %
    {
    The query performance is therefore highly impacted by the speed of the CRC implementation\footnote{We use the generic Boost libraries' CRC: \url{https://www.boost.org/}.}, and more optimized implementations should see a performance increase.

    \nopagebreak
    \boldskipper
    \noindent\textbf{Takeaway:} Because of RDMA, our Key-Value store \emph{can insert entries faster than the CPU can query}. The performance of the CRC implementation plays a key role.
    }

    \subsubsection{Redundancy Effectiveness}
        \label{sec:eval_keywrite_redundancy_effectiveness}
        \begin{figure}[]
            \centering
            \includegraphics[width=\columnwidth]{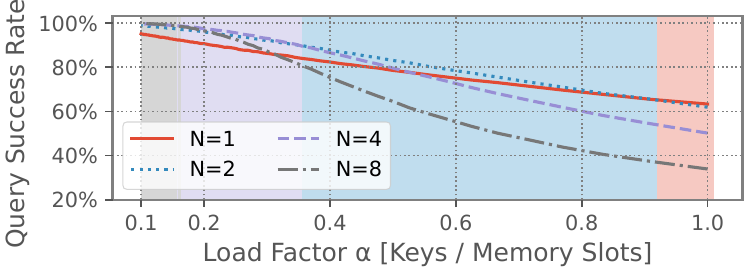}
            \caption{Average query success rates delivered by the Key-Write primitive, depending on the key-value store load factor and the number of addresses per key ($N$). The background color indicates optimal $N$ in each interval.}
            \label{fig:success_rate_N}
        \end{figure}
        
        The probabilistic nature of Key-Write cannot guarantee final queryability on a given reported key due to hash collisions with newer data entries.
        We show in Figure~\ref{fig:success_rate_N} how the query success rate\footnote{The query success rate is defined as the probability at which a previously reported key can be queried from the key-value store.} depends on the load factor (i.e., the total number of telemetry keys over available memory addresses), and the redundancy level ($N$). There is a clear data resiliency improvement by having keys write to $N>1$ memory addresses when the storage load factor is in reasonable intervals. When the load factor increases, adopting more addresses per key does not help because it is harder to reach consensus at query time. The background color in Figure~\ref{fig:success_rate_N} indicate which $N$ delivered the highest key-queryability in each interval.
        
        Higher levels of redundancy improve data longevity, but at the cost of reduced collection and query performance as demonstrated previously in Figures~\ref{fig:eval_keywrite_rate}~and~\ref{fig:eval_keywrite_querying}. Determining an optimal redundancy level therefore has to be a balance between an enhanced data queryability and a reduction in primitive performance, and $N=2$ is a generally good compromise, showing great queryability improvements over $N=1$.

        \nopagebreak
        \boldskipper
        \noindent\textbf{Takeaway:} Increasing the redundancy of all keys does not always improve the query success rate. An optimal redundancy should be set on a case-by-case basis.

    \subsubsection{Data Longevity}
    \label{subsub:longevity}
    
        \begin{figure}[]
            \centering
            \includegraphics[width=\columnwidth]{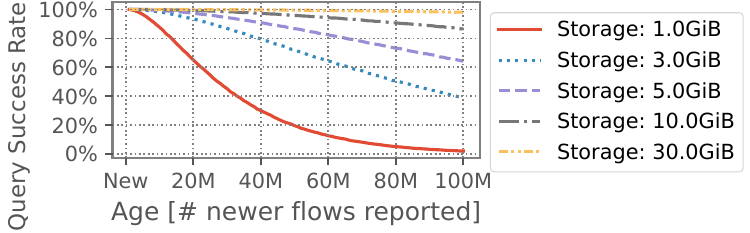}  
            \caption{DTA Key-Write ages out eventually. This figure shows INT 5-hop path tracing queryability of 100 million flows at various storage sizes.}
            \label{fig:eval_key_write_aging}
        \end{figure}
        Data reported by the Key-Write primitive will age out of memory over time due to hash collisions with subsequent reports, which overwrites the memory slots. Figure~\ref{fig:eval_key_write_aging} shows the queryability of randomly reported INT 5-hop path tracing data (i.e., $20B$) at various storage sizes and report ages, with redundancy level $N=2$ and $4B$ checksums. 
        For example, a key-value storage as small as 3GiB is enough to deliver $99.3\%$ successful queries against flows with as many as 10 million subsequently reported paths, which however falls to $44.5\%$ when 100 million subsequent flow are stored in the structure.
        However, increasing storage to 30GiB would allow an impressive $99.99\%$ query success rate for paths with 10 million subsequent reports, or $98.2\%$ success even for flows as old as 100 million subsequent reports.

        \nopagebreak
        \boldskipper
        \noindent\textbf{Takeaway:} It is possible to record data from around $10$M flows in the key-value store while maintaining a $99.99\%$ queryability with just 30GiB of storage.
    
    \subsection{Postcarding Primitive Performance}
    \label{sub:postcard-perf}
    \vspace{1mm}

\begin{figure}
            \centering
            \includegraphics[width=\columnwidth]{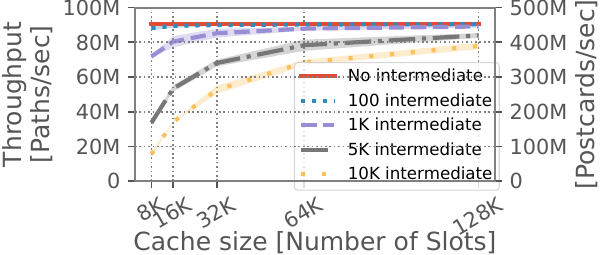}
            \caption{INT-XD/MX postcard collection, using the DTA Postcarding primitive. A report is defined as a successfully aggregated 5-hop path (containing 5 postcards, one per hop).}
            \label{fig:eval_postcarding_rate}
        \end{figure}

        The Postcarding primitive has been benchmarked for aggregating and collecting INT-XD/MX postcards across 5-hop network paths.
        The number of other flows appearing at the translator while aggregating per-flow postcards increases the risk of premature cache emission.
        Figure~\ref{fig:eval_postcarding_rate} shows us the effect that the number of intermediate flows and the size of the cache has on the aggregation performance, with a maximum achieved collection rate of $90.5\mathit{MPaths/s}$ ($452.5\mathit{MPostcards/s}$).\footnote{Early emissions (i.e., path-reports with missing postcards) are counted as failures in this test despite being potentially useful (e.g., knowing 4 out of 5 hops in a path), and are not included in the collection throughput.
        }
        Comparing the performance to Key-Write in Figure~\ref{fig:eval_keywrite_rate}, where we would need $5$ different reports to collect a full path, we see a significant performance gain \mbox{by the Postcarding primitive.}

        \nopagebreak
        \boldskipper
        \noindent\textbf{Takeaway:} The performance of Postcarding depends on the rate of cache collisions in the translator during the aggregation-phase, and can \emph{improve upon the best-case Key-Write performance by up to $4.3x$} for 5-hop collection.

    \subsection{Append Primitive Performance}
    \label{sec:eval_append_performance}

    \begin{figure}
        \centering
        \includegraphics[width=\columnwidth]{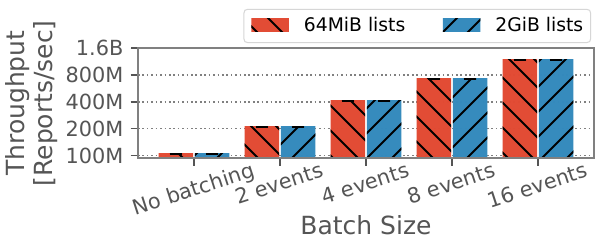}
        \caption{Telemery event-report collection, using DTA Append and different batch sizes.
        Performance increases linearly with batch sizes until we achieve line-rate with batches of $4x4B$. {The collection speed is not impacted by the list sizes.}}
        \label{fig:eval_append_rate}
    \end{figure}
            
    We have benchmarked the performance of the Append primitive for collecting telemetry event-reports, both at different batch sizes and total size of the allocated data list, while reporting data into a single \mbox{list. The results are shown in Figure~\ref{fig:eval_append_rate}.}
    
    We noticed no performance impact from different report sizes, until we reached the line-rate of $100G$ for large batch sizes after which the performance increased sub-linearly. The results in Figure~\ref{fig:eval_append_rate} show this effect for $4B$ queue-depth reports, where we reach line-rate at batches of $4$. Our base performance is bounded by the RDMA message rate of the NIC, which is the current collection bottleneck in our system, and the high performance of the Append primitive is due to including several reports in each memory operation. Performing equivalent tests with up to $131K$ parallel lists showed a negligible performance impact.
    
    \nopagebreak
    \boldskipper
    \noindent\textbf{Takeaway:} The Append primitive is able to collect \emph{over $1$ billion telemetry event reports per second}.
            
    \subsubsection{Append List-Polling Rate}
    \label{sec:eval_append_querying}
    
    \begin{figure}
        \centering
        \begin{subfigure}[b]{0.667\columnwidth}
            \centering
            \includegraphics[width=\linewidth]{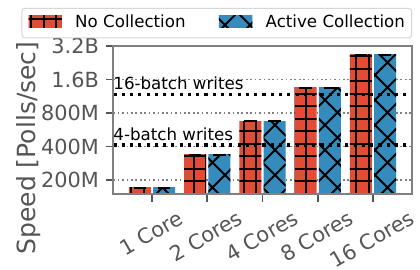}
            \caption{Polling rate.}
            \label{fig:eval_append_query_rate}
        \end{subfigure}
        \hfill
        \begin{subfigure}[b]{0.310\columnwidth}
            \centering
            \includegraphics[width=\linewidth]{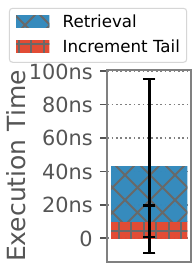}
            \caption{Poll breakdown.}
            \label{fig:eval_append_querying_breakdown}
        \end{subfigure}
        \caption{Append primitive querying performance. Append-lists are queried either while collecting no reports or at $50\%$ capacity (while collecting \emph{600M} reports per second). Collection has a negligible impact on data retrieval rate, and processing rate scales near-linearly with the number of cores. The dotted lines show \mbox{the maximum collection rates at different batch sizes.}}
        \label{fig:eval_append_querying}
    \end{figure}
                
    Figure~\ref{fig:eval_append_query_rate} shows the raw list polling rates, which is the speed at which appended data can be read into the CPU for processing. We assume that collection runs simultaneously to the CPU reading data from the lists, by having the translator process $600$ million Append operations per second in batches of size 16, which approximates collection at half capacity. Simultaneously collecting and processing telemetry data show no noticeable impact on either collection or processing, showing that DTA is not memory-bounded even at this speed\footnote{DTA is neither memory- nor CPU bounded in these tests, regardless of the collection rate, but is instead limited by the message rate of the network card}.
                
    Extracting telemetry data from the lists is a very lightweight process, as shown in Figure~\ref{fig:eval_append_querying_breakdown}, requiring a pointer increment, possibly rolling back to the start of the buffer, and then reading the memory location.
    We allocated a number of lists equal to the number of CPU cores used during the test to prevent race conditions at the tail pointer.
    %
    Our tests showed that just $8$ cores proved capable of extracting every telemetry report even when large batches reported at maximum capacity. This leaves us much processing power for complex real-time telemetry processing.
    We see that the collector \mbox{can even retrieve list entries faster than the RAM clock speed.} 

    \nopagebreak
    \boldskipper
    \noindent\textbf{Takeaway:} The CPU retrieves appended reports \emph{faster than they can be collected} \mbox{(Figure~\ref{fig:eval_append_rate}), with margin left for further processing.}
    
    \vspace{-1mm}

\section{Discussion}
    \label{sec:discussion}

    \boldskipper
    \noindent\textbf{The generality and scope of DTA.}
    DTA is not intended to be a competitor of existing data plane assisted monitoring systems~\cite{INTSpec,kim2015band,narayana2017language,gupta2018sonata,ben2020pint,teixeira2020packetscope,zhou2020flow,teixeira2020packetscope,sonchack2018turboflow}. These either focus on extracting new metrics or reducing the costs of telemetry monitoring through intelligent pre-processing and filtering within the switching ASIC. 
    Nevertheless, these systems generate a significant amount of telemetry information, especially with large-scale networks, multiple queries, and/or fine telemetry granularities (Table~\ref{tab:motivation_generation_speeds}).
    
    DTA can be coupled with existing telemetry systems and serve as an interface between the on-switch monitoring functions and the telemetry analysis back-end in the control plane. To achieve broad compatibility with a variety of monitoring solutions, we have designed several generic and highly flexible primitives to simplify the integration of DTA into both existing and future telemetry environments. As a consequence, with DTA, we replace only the report ingestion mechanism of the telemetry collector (e.g., DPDK along with data structure population), not the rest of the collector (e.g., data analysis and decision-making). For example, it is possible to couple the streaming analysis engine of Sonata~\cite{gupta2018sonata} with DTA: in this scenario our solution is in charge of transferring data from switches to collector's memory, while the original Sonata's engine performs analysis on the received data. For a more extensive list of examples, we refer to Table~\ref{tab:primitive_mappings} that recap how DTA can be integrated into various telemetry systems to enhance their performances.
    
    \noindent\textbf{Implementing the translator in a SmartNIC.} There are two main approaches we have considered on where to deploy the translator: a SmartNIC located at the collector and the last-hop programmable switch (which we explored in this work). A SmartNIC would allow us to completely remove RDMA traffic: the NIC dataplane would process incoming DTA packets and translate them into local DMA calls. Exploring DTA translation in SmartNICs is left for future work. Nevertheless, we believe that our P4 implementation can be a starting point for P4-capable NICs~\cite{pensandoNICs}.

    \boldskipper
    \noindent\textbf{Supporting Multiple Collectors.} 
    It is beneficial to enable collection at multiple servers for scalability or resiliency. DTA can be deployed alongside multiple collectors and permit easy partitioning of reports based on the IP and DTA headers.

    \boldskipper
    \noindent\textbf{Flow Control in DTA.}
    Best-effort transport protocols, e.g., UDP, are used by many well-known telemetry systems (e.g.,~\cite{ciscoNetflow,intelINT}).
    Similarly, DTA does not assure reliable delivery. However, it can be used in conjunction with flow control mechanisms that allow for lossless \mbox{delivery of data~\cite{pfc,bfc}. }
    

    \boldskipper
    \noindent\textbf{Query-Enhancing Extensions.}
    In some cases, queries may be\linebreak known ahead of time, in which case our translator can aid in their processing.
    For example, while switches can measure the queuing latency of a flow, we are often interested in knowing the end to end delay~\cite{rtt-sigcomm}, which can be expressed as follows:
        \begin{lstlisting}[
           language=SQL,
           showspaces=false,
           basicstyle=\ttfamily,
           numbers=none,%left
           numberstyle=\tiny,
           commentstyle=\color{gray}
        ]
SELECT flowID,path WHERE SUM(latency) > T
        \end{lstlisting}
        Knowing the query ahead of time, our translator can wait for postcards from all switches through which the SYN packet of the flow was routed, sum their latency, and \mbox{report it if it is over the threshold.}
    
    \boldskipper
    \noindent\textbf{Push notifications.} An advantage CPU-based collectors have over DTA is that the CPU can trigger analysis tasks as soon as it receives reports.
    In our case, for key-value store operations, the CPU must first find out if new data has been written into the memory;  however, we assume for Append operations the CPU is monitoring the lists continuously, which would allow for equivalent reactivity to CPU-based solutions. Additionally, DTA packets can include an \emph{immediate flag}, which can be used by the translator to inform the CPU that new data has arrived through RDMA immediate interrupts (e.g., a flow is experiencing problems). Deciding which reports should carry such a flag is beyond the scope of this work.

    \boldskipper
    \noindent\textbf{The next telemetry bottleneck.} DTA significantly reduces the cost of telemetry ingestion mainly by bypassing any CPU processing. In our experiments  the new bottleneck is the message rate of the RDMA NICs at the collectors. 
    To address this message rate limitation, DTA already supports multi-NIC collectors. Future NICs  will have better speeds.

    A possible future bottleneck is the memory speed where we store the telemetry data structures. However, current-generation DRAM can achieve billions of memory transfers per second and is likely to increase further in the future.
    Therefore, it is possible that telemetry ingestion itself might no longer be seen as the main bottleneck in telemetry systems going forward, if the CPU is bypassed. 
    Instead, given the increasing sophistication and complexity of data analysis tools, the de-facto bottleneck might instead be the rate at which we can still meaningfully analyze the generated data in real time.

    
\section{Related Work}
    \vspace*{1mm}

    \noindent\textbf{Telemetry and Collection.} Traditional techniques for monitoring the status of the network have looked into periodically collecting telemetry data~\cite{guo2015pingmesh,hare2011simple} or mirroring packets at switches~\cite{zhu2015packet,planck}. The former generates coarse-grained data that can be significant given the large scale of today's networks~\cite{stroboscope}. The latter has been recognized as viable option only if it is known in advance the specific flow to monitor~\cite{zhu2015packet}. The rise in programmable switches has enabled fine-grained telemetry techniques that generate a lot more data~\cite{tammana2018distributed,INTSpec,ben2020pint,zhou2020flow,zhu2015packet, gupta2018sonata}.
    Irrespective of the techniques, collection is identified to be the main bottleneck in network-wide telemetry, and previous works focus on either optimizing the collector stack performance~\cite{khandelwal2019confluo,van2018intcollector}, or reducing the load through offloaded pre-processing~\cite{li2020concerto} and in-network filtering~\cite{intelINT,elastictrie,vestin2019programmable,zhou2020flow}.
    In an earlier version of our project, we investigated the possibility of entirely bypassing collectors' CPU, but limited the collection process to data that can be represented as a key-value store~\cite{langlet2021zero}. DTA expands it significantly by introducing the translator, designing additional primitives, building a prototype, investigating the systems aspects, and showing an end-to-end improvement over state-of-the-art collection systems.
    In particular, this paper proposes an alternative solution which is generic and works with a number of existing state-of-the-art monitoring systems.
    We show examples of where these aforementioned systems can integrate DTA earlier in Table~\ref{tab:primitive_mappings}.
    A further alternative approach is letting the end-hosts assist in network-wide telemetry~\cite{tammana2018distributed,huang2020omnimon}, which unfortunately requires significant investments and infrastructure changes and still lean on centralized collection to achieve a network-wide view.

    \boldskipper
    \noindent\textbf{RDMA in programmable networks.} Recent works have shown that programmable switches can perform RDMA operations to access server DRAM for expanded memory in their stateful network functions~\cite{kim2020tea,scazzariello2023high}. These works are interesting for these scenarios, but are not suited for the queryable aggregation required for telemetry collection.
    Programmable network cards are also shown capable of expanding upon RDMA with new and customized primitives~\cite{amaro2020remote}. Especially FPGA network cards show great promise for high-speed custom RDMA verbs~\cite{sidler2020strom,mansour2019fpga}. 
    However, as discussed in Section~\ref{sec:motivation}, telemetry collection brings new challenges when used in conjunction with the RoCEv2 protocol.
    As previously mentioned earlier in discussion, a protocol such as DTA that is tailored for telemetry collection could very well be implemented at the NIC-level. 
    
\section{Conclusion}
\vspace{1mm}
    We presented Direct Telemetry Access (DTA), a new telemetry collection system optimized for storing reports from switches to collectors' memory. We built DTA on top of RDMA and provided novel and expressive primitives that allow easy integration with existing telemetry solutions.
    
    DTA can write to our key-value store over $400$M INT reports per second, without any CPU processing, $16$x better than the state-of-the-art collector.
    When the received data can be recorded sequentially, as in the case of temporally ordered event reports, it can ingest up to a billion reports per second, a $41$x improvement over the state-of-the-art.

 \medskip
    \noindent\textbf{This work does not raise any ethical issues.}
    
    \section*{Acknowledgements}
    We thank our shepherd Kate Lin, and the anonymous reviewers, for valuable comments and feedback.
    This work was supported in part by ACE, one of the seven centers in JUMP 2.0, a Semiconductor Research Corporation (SRC) program sponsored by DARPA, by the UK EPSRC project EP/T007206/1, by the European Union under the Italian National Recovery and Resilience Plan (NRRP) of NextGenerationEU, partnership on “Telecommunications of the Future” (PE00000001 - program “RESTART”), and by a gift from Facebook/Meta.
    Michael Mitzenmacher was supported in part by NSF grants CCF-2101140, CNS-2107078, and DMS-2023528.
    Finally, a big thanks to Sivaram Ramanathan for invaluable input in the early stages of the project.

    \appendix
   
    \bibliographystyle{plain}
    \bibliography{references}

    
\clearpage
\noindent
Appendices are supporting material that has not been peer-reviewed.
\section{Appendix}
    \subsection{Key-Write Algorithm}
        \begin{figure}[h!]
            \centering
            \includegraphics[width=0.7\linewidth]{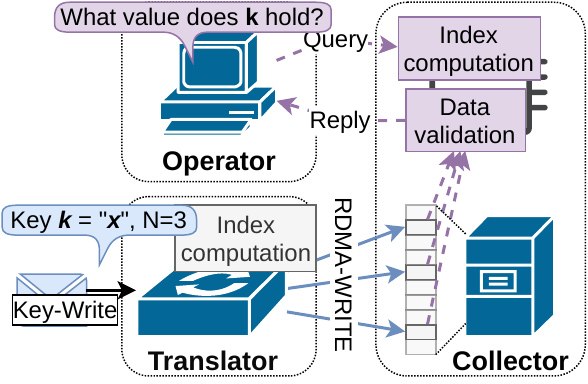}
            \caption{Key-Write Overview.}
            \label{fig:dta_keywrite}
        \end{figure}
        The Key-Write primitive is an abstraction around a key-value store, allowing read/writes of telemetry data. Figure~\ref{fig:dta_keywrite} is a high-level visualization of the primitive, and algorithm pseudo-code is presented in Algorithm~\ref{alg:keywrite_translation}~and~\ref{alg:keywrite_querying}.

        \begin{algorithm}[]
            \DontPrintSemicolon
            \SetKwInOut{Input}{Input}
            \SetKwFunction{Fnc}{CraftWrite}
            \SetKwProg{Fn}{Function}{}{end}
            
            \Input{Redundancy $N$, Key $K$, Telemetry data $D$}
            $\mathit{Bufstart} \gets$ Address to start of RDMA memory buffer\;
            $\mathit{Buflen} \gets$ Number of allocated KeyVal slots\;
            $\mathit{Slotlen} \gets$ Size of one KeyVal slot\;
            
            \Fn{\Fnc{$n$, $K$, $D$}}{
                $\mathit{Slot} \gets h_0(n, K) \bmod \mathit{Buflen}$\;
                $\mathit{Dest} \gets \mathit{Bufstart} + \mathit{Slot} \times \mathit{Slotlen}$\;
                $\mathit{Csum} \gets h_1(K)$\;
                Write $(\mathit{Csum},D)$ to address $\mathit{Dest}$ through RDMA\;
            }
            
            \For{$n = 0$ \KwTo $N$}{
                \Fnc{$n$, $K$, $D$}\;
            }
            
            \caption{DTA-to-RDMA translation in Key-Write}
            \label{alg:keywrite_translation}
        \end{algorithm}

        \begin{algorithm}[]
            \DontPrintSemicolon
            \SetKwInOut{Input}{Input}
            \SetKwInOut{Output}{Output}
            \SetKwFunction{Fnc}{GetSlot}
            \SetKwProg{Fn}{Function}{}{end}
            
            \Input{Redundancy $N$, Key $K$, Consensus threshold $T$}
            \Output{$\mathit{D_{winner}}$}
            $\mathit{Buflen} \gets$ Number of allocated KeyVal slots\;
            $\mathit{Storage} \gets$ Array size $\mathit{Buflen}$ with $\langle \mathit{Csum}, D \rangle$ elements\;
            
            \Fn{\Fnc{$n$, $K$}}{
                $\mathit{Slot} \gets h_0(n, K) \bmod \mathit{Buflen}$\;
                \Return $\mathit{Storage}[\mathit{Slot}]$\;
            }
            
            $\mathit{Csum} \gets h_1(K)$\;
            
            \For{$n = 0$ \KwTo $N$}{
                $(\mathit{Csum_{slot}},D) \gets$ \Fnc{$n$, $K$}\;
                \If{$\mathit{Csum} == \mathit{Csum_{slot}}$}{
                    Add $D$ to list of candidates\;
                }
            }
            
            $\mathit{D_{winner}} \gets$ candidate $D$ if $D$ appears at least $T$ times\;
            
            \caption{Querying the Key-Write storage}
            \label{alg:keywrite_querying}
        \end{algorithm}

    \newpage
    \subsection{Postcarding Algorithm}
        \begin{figure}[h!]
            \centering
            \includegraphics[width=0.7\linewidth]{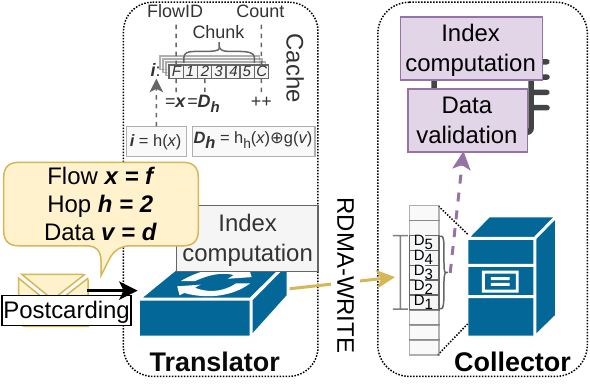}
            \caption{Postcarding Overview.}
            \label{fig:dta_postcarding}
        \end{figure}
        
        The Postcarding primitive is an abstraction around a key-value store with per-flow aggregation of INT postcards, allowing read/writes of telemetry data. Figure~\ref{fig:dta_postcarding} is a high-level visualization of the primitive. We refer to Section~\ref{sec:design} for details on primitive translation and querying, as well as Appendix~\ref{app:postcarding} for analysis.

    \newpage
    \subsection{Append Algorithm}
        \begin{figure}[h!]
            \centering
            \includegraphics[width=0.7\linewidth]{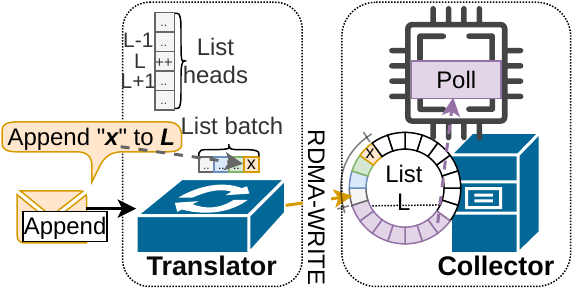}
            \caption{Append Overview.}
            \label{fig:dta_append}
        \end{figure}
        
        The Append primitive is an abstraction around data lists, allowing read/insertion of telemetry data. Figure~\ref{fig:dta_append} is a high-level visualization of the primitive, and algorithm pseudo-code is presented in Algorithm~\ref{alg:append_translation}~and~\ref{alg:append_querying}.

        \begin{algorithm}[]
            \DontPrintSemicolon
            \SetKwInOut{Input}{Input}
            \SetKwFunction{Fnc}{WriteBatch}
            \SetKwProg{Fn}{Function}{}{end}
            
            \Input{List ID $L$, Data $D$}
            $\mathit{ListBuffers} \gets$ Vector with $|\mathit{Lists}|$ buffer pointers\;
            $\mathit{BufferLengths} \gets$ Vector with $|\mathit{Lists}|$ buffer lengths\;
            $\mathit{Heads} \gets$ Vector with $|\mathit{Lists}|$ head-offsets\;
            $\mathit{BatchSize} \gets$ The global batch size\;
            $\mathit{BatchPointer} \gets$ Vector with $|\mathit{Lists}|$ integers\;
            $\mathit{Batches} \gets$ 2D-vector sized $[|\mathit{Lists}|][\mathit{BatchSize}-1]$\;
            
            \Fn{\Fnc{L,D}}{
                $\mathit{Batch} \gets (\mathit{Batches}[L], D)$\;
                $\mathit{Address} \gets \mathit{ListBuffers}[L] + \mathit{Heads}[L]$\;
                Write $\mathit{Batch}$ to address $\mathit{Address}$ through RDMA\;
                $\mathit{Heads}[L] \mathrel{+}= \mathit{BatchSize}$\;
                \If{$\mathit{Heads}[L] == \mathit{BatchSize}$}{
                    $\mathit{Heads}[L] \gets 0$\;
                }
            }
            
            \If{$\mathit{BatchPointer}[L] == \mathit{BatchSize}$}{
                \Fnc{L,D}\;
                $\mathit{BatchPointer}[L] \gets 0$\;
            }
            \Else{
                $\mathit{Batches}[L][\mathit{BatchPointer}[L]] \gets D$\;
                $\mathit{BatchPointer}[L] \plusplus$\;
            }
            
            \caption{DTA-to-RDMA translation in Append}
            \label{alg:append_translation}
        \end{algorithm}

        \begin{algorithm}[]
            \DontPrintSemicolon
            \SetKwInOut{Input}{Input}
            \SetKwInOut{Output}{Output}
            
            \Input{List ID $L$}
            \Output{$\mathit{data}$}
            $\mathit{ListBuffers} \gets$ Vector with $|\mathit{Lists}|$ buffer pointers\;
            $\mathit{BufferLengths} \gets$ Vector with $|\mathit{Lists}|$ buffer lengths\;
            $\mathit{Heads} \gets$ Vector with $|\mathit{Lists}|$ head-offsets\;
            $\mathit{data} \gets \mathit{ListBuffers}[L] + \mathit{Heads}[L]$\;
            $\mathit{Heads}[L] \gets (\mathit{Heads}[L] + 1) \bmod \mathit{BufferLengths}[L]$\;
            
            \caption{Querying the Append storage}
            \label{alg:append_querying}
        \end{algorithm}

    \newpage
    \subsection{Key-Increment Algorithm}
        \begin{figure}[h!]
            \centering
            \includegraphics[width=0.7\linewidth]{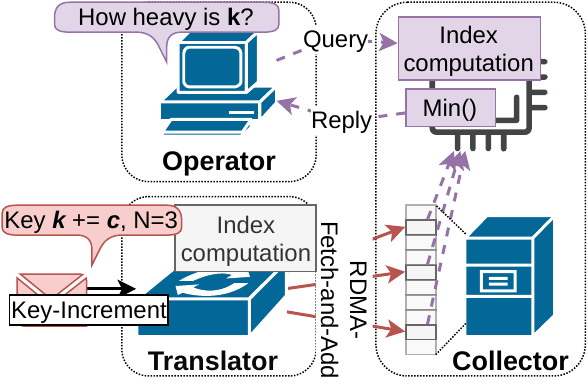}
            \caption{Key-Increment Overview.}
            \label{fig:dta_keyincrement}
        \end{figure}

        The Key-Increment primitive is an abstraction around a key-value store, allowing read/increment of counters. Figure~\ref{fig:dta_keyincrement} is a high-level visualization of the primitive, and algorithm pseudo-code is presented in Algorithm~\ref{alg:keyincrement_translation}~and~\ref{alg:keyincrement_querying}.

        \begin{algorithm}[]
            \DontPrintSemicolon
            \SetKwInOut{Input}{Input}
            \SetKwFunction{Fnc}{CraftWrite}
            \SetKwProg{Fn}{Function}{}{end}
            
            \Input{Redundancy $N$, Key $K$, Counter $C$}
            $\mathit{Bufstart} \gets$ Address to start of RDMA memory buffer\;
            $\mathit{Buflen} \gets$ Number of allocated KeyVal slots\;
            
            \Fn{\Fnc{$n,K,C$}}{
                $\mathit{Slot} \gets h_0(n,K) \bmod \mathit{Buflen}$\;
                $\mathit{Dest} \gets \mathit{Bufstart} + \mathit{Slot} * 4$\;
                Increment $\mathit{Dest}$ by $\mathit{C}$ through RDMA Fetch\&Add\;
            }
            
            \For{$n = 0 \rightarrow N$}{
                \Fnc{$n,K,C$}\;
            }
            
            \caption{DTA-to-RDMA translation in Key-Increment}
            \label{alg:keyincrement_translation}
        \end{algorithm}

        \begin{algorithm}[]
            \DontPrintSemicolon
            \SetKwInOut{Input}{Input}
            \SetKwInOut{Output}{Output}
            \SetKwFunction{Fnc}{GetSlot}
            \SetKwProg{Fn}{Function}{}{end}
            
            \Input{Redundancy $N$, Key $K$}
            \Output{$\mathit{C_{winner}}$}
            $\mathit{Buflen} \gets$ Number of allocated KeyVal slots\;
            $\mathit{Storage} \gets$ Array size $\mathit{Buflen}$ with $\langle C \rangle$ elements\;
            
            \Fn{\Fnc{$n,K$}}{
                $\mathit{Slot} \gets h_0(n,K) \bmod \mathit{Buflen}$\;
                \Return $\mathit{Storage}[\mathit{Slot}]$\;
            }
            
            $\mathit{Counters} \gets \lbrack \rbrack$\;
            \For{$n = 0 \rightarrow N$}{
                $\mathit{Counters}[n] \gets \Fnc{n,K}$\;
            }
            $\mathit{C_{winner}} \gets \min(\mathit{Counters})$\;
            
            \caption{Querying the Key-Increment storage}
            \label{alg:keyincrement_querying}
        \end{algorithm}

\newpage
\subsection{Analysis of the Key-Write primitive}            
\vspace*{-1mm}
\label{sec:kw-theory}
Because we treat the RDMA memory as a large key-value hash table where only checksums of keys are stored and values may be overwritten over time, we must consider the possibility that when we make a query, we are unable to return an answer, or we may return an incorrect answer.  
We call the case where we have no answer to return an {\em empty return}, and the case where we return an incorrect answer a {\em return error}.
The probability of an empty return or a return error depends on the parameters of the system, and on the method we choose to determine the return value.  Below we present some of the possible tradeoffs and some mathematical analysis; we leave \mbox{further results and discussions for the full paper.}

Let us first consider a simple example.  When a write occurs for a key-value pair, in the hash table $N$ copies of the $b$-bit key checksum and the value are stored at random locations. 
We assume the checksum is uniformly distributed for any given key throughout our analysis.
When a read occurs, let us suppose we return a value if there is only a single value amongst the $N$ memory locations matching that checksum.  (The value could occur multiple times, of course.)

An empty return can occur, for example, if when we search the $N$ locations for a key, none of them have the right checksum.  
That is, all $N$ copies of the key have been overwritten, and none of the $N$ locations currently hold another key with the same checksum.
To analyze this case, let us consider the following scenario.  
Suppose that we have $M$ memory cells total, and that there are $K = \alpha M$ updates of {\em distinct} keys between when our query key $q$ was last written, and when we are making a query for its values.  
We can use the Poisson approximation for the binomial (as is standard in these types of analyses and accurate for even reasonably large $M$, $N$, $K$; see, for example, \cite{broder2004network,mitzenmacher2017probability}).
Using such approximations, the probability that any one of the $N$ locations is overwritten is given by $(1-e^{-KN/M})$, and that all of them are overwritten is $(1-e^{-KN/M})^N$.  
The probability that all of them are
overwritten and \mbox{the key checksum is not found is approximated by} $$(1-e^{-KN/M})^N \cdot (1-2^{-b})^N = (1-e^{-\alpha N})^N \cdot (1-2^{-b})^N.$$

We would also get an empty return if the $N$ cells contained two or more distinct  values with the same correct checksum.

\vspace{1mm}\noindent
This probability is lower bounded by
{\small{
$$ \sum_{j=1}^{N-1} {N \choose j} (1-e^{-\alpha N })^j e^{-\alpha N(N-j)} (1 - (1-2^{-b})^j)\quad ,$$ 
}}
and upper bounded by  
{\small{
\begin{multline*}
    \Big( \sum_{j=1}^{N-1} {N \choose j} (1-e^{-\alpha N})^j e^{-\alpha N(N-j)} (1 - (1-2^{-b})^j) \Big)\\
    \qquad\qquad + (1-e^{- \alpha N})^N (1 - (1-2^{-b})^N - N\cdot 2^{-b}(1-2^{-b})^{N-1}).
\end{multline*}
}} 
\noindent The first summation is the probability at least one of the original $N$ locations is not overwritten, but at least one overwritten location gets the same checksum.  (We pessimistically assume it obtains a different value.)
The second expression adds a term for when all original values are overwritten and two or more obtain the same checksum. Note that we need to give bounds as values in overwritten locations \mbox{may or may not be the same.}

We could have a return error if all $N$ copies of the original key are overwritten and one or more of those cells are overwritten with the same checksum and same (incorrect) value.  
This probability is lower bounded by $$(1-e^{-\alpha N})^N N 2^{-b} (1-2^{-b})^{N-1},$$  which is the probability that all of the original locations are overwritten and a single overwriting key obtains the checksum, and upper bounded by   $$(1-e^{-\alpha N})^N (1 - (1- 2^{-b})^N),$$  the probability that the original locations are overwritten and at least one overwriting key obtains the checksum.

There are many ways to modify the configuration or return method to lower the empty returns and/or return errors, at the cost of more computation and/or more memory.  
The most natural is to simply use a larger checksum; we suggest 32 bits should be appropriate for many situations. 
However, we note that at ``Internet scale'' rare events will occur, even matching of 32-bit checksums, and so this should be considered when utilizing Key-Write information.
One can also use a ``plurality vote'' if more than one value appears for the queried checksum;  additionally one can require that a checksum/value pair occur at least twice among the $N$ values before being returned.  (Note that, for example, requiring consensus of two values can be decided on a per query basis without changing anything else; one can decide for specific queries whether to trade off empty returns and return errors this way.) 
Additional ideas from coding theory \cite{goodrich2011invertible,lipton1994new}, including using different checksums for each location or XORing each value with a pseudorandom value, could also be applied.  
As a default, we suggest a 32-bit checksum and a ``plurality vote''. 

\clearpage
\subsection{Analysis of the Postcarding Primitive}\label{app:postcarding}
    We now calculate:\begin{enumerate}[label=(\alph*)]
        \item The probability that a flow's values fail to be reported, because the flow has been overwritten.\label{postcards_L1}
        \item The probability that a flow is reported with incorrect values.\label{postcards_L2}
    \end{enumerate}
    We assume that the number of \emph{reports} (up to $B$ postcards that belong to the same flow/packet) since the queried ID is $\alpha\cdot C$.
    
    For~\ref{postcards_L1}, we consider several reasons (similar to~\eqref{eq1}-\eqref{eq3}) for failing to report the values and analyze them separately.

    \begin{itemize}
        \item All of the queried flow's chunks are overwritten by other flows and none of them produce valid information.
        We have that the probability that a slot is overwritten is bounded by
        $
            (1-e^{-\alpha\cdot N}). 
        $
        Also, the probability of a given overwritten slot to \textit{not} produce valid information is:
        $
            1-\parentheses{(|V|+1)\cdot2^{-b}}^B.
        $
        Therefore, the overall probability of this event is at most
        \begin{align}
            (1-e^{-\alpha\cdot N})^N\cdot \parentheses{1-\parentheses{(|V|+1)\cdot2^{-b}}^B}^{N}.\label{eq5}
        \end{align}
        \item All the flow's chunks are overwritten and at least two produce valid information arrays that differ. This probability is bounded by:
        \begin{multline}
        \hspace{-0mm}
            (1-e^{-\alpha\cdot N})^N\cdot\Bigg(1-\parentheses{1-\parentheses{(|V|+1)\cdot2^{-b}}^B}^{N}\\\hspace{-0mm}\qquad- N\cdot \parentheses{(|V|+1)\cdot2^{-b}}^B\\\hspace{-0mm}\qquad\cdot\parentheses{1-\parentheses{(|V|+1)\cdot2^{-b}}^B}^{N-1}\Bigg).\label{eq6}
        \end{multline}
        \item At least one chunk (but not all) is overwritten and produces valid information. This error probability is at most
        {\small
        \begin{multline}
            \sum_{j=1}^{N-1} {N \choose j}\cdot (1-e^{-\alpha\cdot N})^j\cdot e^{-\alpha\cdot N(N-j)}\\\cdot \parentheses{1-\parentheses{1-\parentheses{(|V|+1)\cdot2^{-b}}^B}^{j}}.\label{eq7}
        \end{multline}
        }
        %
    \end{itemize}
    Next, we analyze the probability of replying incorrectly~\ref{postcards_L2}. This happens when all the queried key's chunks are overwritten and all valid chunks are hold the same information. Thus, the probability of such an error is at most:
    \begin{equation}
        (1-e^{-\alpha\cdot N})^N\cdot  N\cdot \parentheses{ (|V|+1)\cdot2^{-b}}^ B
        .\label{eq8}
    \end{equation}
    
    Let us consider a numeric example to contrast these results with using KW for each report of a given packet. Specifically, suppose that we are in a large data center ($|V|=2^{18}$ switches) and want to run path tracing by collecting all (up to $B=5$) switch IDs using $N=2$ redundancy.
    Further, let us set $b=32$-bit per report and compare it with $64$ bits ($32$ for the key's checksum and $32$ bits for the switch ID) used in KW and that $C\cdot \alpha$ packets' reports were collected after the queried one, for $\alpha=0.1$.
    We have that the probability of not outputting a collected report (\ref{eq5}-\ref{eq7}) is at most 3.3\% and the chance of providing the wrong output \eqref{eq8} is lower than $10^{-22}$. In contrast, using KW for postcarding gives a false output probability of $\approx 8\cdot 10^{-11}$ (in at least one hop) using twice the width per entry! This improvement is due to a couple of reasons. First, we leverage the difference between the number of switches (e.g., $|V|=2^{18}$) and the width of the value field (hardcoded at 32-bits per the INT standard~\cite{INTSpec}). Second, we leverage the fact that each packet carries multiple (e.g., $B=5$) reports to amplify the success probability and mitigate the chance of wrong output. Further, for reports for which we are able to cache all postcards at the translator (which depends on the allocated memory and the number of simultaneous postcard reports generated), this approach reduces the number of RDMA writes by a factor of $B$.

\end{document}